\documentclass[12pt]{article}
\usepackage{cancel}
\usepackage{verbatim}
\usepackage{graphicx}
\usepackage{bm}
\usepackage{array}
\usepackage{epsfig}
\usepackage{makecell}
\usepackage{graphicx}
\usepackage{booktabs}
\usepackage{multirow}
\usepackage{rotating}
\usepackage{lscape}
\usepackage{natbib}
\usepackage{enumitem}
\usepackage{hyperref}
\usepackage{amssymb}
\usepackage{amsthm}
\usepackage{color,xcolor}
\usepackage[cmex10]{amsmath}
\usepackage{subcaption}
\usepackage{threeparttable} 
\usepackage{tablefootnote}
\allowdisplaybreaks

\def\standard{\oddsidemargin=0in
	\evensidemargin=0in
	\topmargin =-.5in
	\textheight=8.4in
	\textwidth=6.6in}
	{\end{list}} 
	{\end{list}} 
\newcounter{bean}
	{\end{list}} 

\makeatletter
\def\equationarray{\stepcounter{equation}
	\let\@currentlabel=\theequation \global\@eqnswtrue
	\global\@eqcnt\z@ \global\@eqargcnt \z@
	\let\\=\@equationcr \let\@acol\@arrayacol
	\let\@classz\@eqnclassz
	\def\@halignto{to\displaywidth}
	\@equationarray}

\def\@equationarray#1{\setbox\@arstrutbox=\hbox{\vrule
		height\arraystretch \ht\strutbox
		depth\arraystretch \dp\strutbox
		width\z@}
	\@mkpream{#1}
	\edef\@preamble{\halign \noexpand\@halignto
		\bgroup \tabskip\z@ \@arstrut \@preamble
		\hfill\tabskip\@centering &\llap{\@sharp}
		\tabskip\z@\cr}%
	\let\@startpbox\@@startpbox\let\@endpbox
	\@@endpbox \bgroup \let\par\relax
	\let\@sharp## \let\protect\relax \lineskip\z@
	\baselineskip\z@
	\tabskip\@centering $$ 
	\@preamble}

\def\@eqnclassz{\ifcase \@lastchclass \@acolampacol\or
	\@ampacol \or \or \or \@addamp \or
	\@acolampacol \or \@firstampfalse \@acol \fi
	\global\advance\@eqargcnt\@ne
	\edef\@preamble{\@preamble
		\ifcase \@chnum
		\hfil$\relax\displaystyle\tabskip\z@\@sharp$\hfil
		\or
		$\relax\displaystyle\tabskip\z@\@sharp$\hfil \or
		\hfil$\relax\displaystyle\tabskip\z@\@sharp$\fi
		\global\advance\@eqcnt\@ne}}

\def\endequationarray{\@equationcr
	\egroup\global\advance\c@equation\m@ne %
	$$\tabskip\@centering\egroup\global\@ignoretrue}

\def\@equationcr{${\ifnum0=`}\fi\@ifstar{\@xequationcr}
	{\@xequationcr}}
\def\@xequationcr{
	\@ifnextchar[{\@argequationcr}{\ifnum0=`{\fi}${}
		\@zequationcr
		\global\@eqnswtrue\global\@eqcnt\z@\cr}}

\def\@argequationcr[#1]{\ifnum0=`{\fi}${}\ifdim #1>\z@
	\@xargequationcr{#1}\else
	\@yargequationcr{#1}\fi}

\def\@xargequationcr#1{
	\@tempdima #1\advance\@tempdima \dp \@arstrutbox
	\vrule \@height\z@ \@depth\@tempdima \@width\z@
	\@zequationcr
	\global\@eqnswtrue\global\@eqcnt\z@ \cr}

\def\@yargequationcr#1{ \@zequationcr
	\global\@eqnswtrue\global\@eqcnt\z@
	\cr\noalign{\vskip #1}}

\def\@zequationcr{\@wargequationcr\if@eqnsw\@eqnnum
	\stepcounter{equation}\global\@eqcnt\z@\fi}

\def\@wargequationcr{\ifnum\@eqcnt<\@eqargcnt
	\@amper \@wargequationcr\fi}

\def\@amper{&}

\newcount\@eqargcnt

\makeatother 

\def\b1{{\bf 1}}

\def\qed{\quad \mbox{$\vcenter{\vbox{\hrule height.4pt
				\hbox{\vrule width.4pt height.9ex \kern.9ex \vrule
					width.4pt} \hrule height.4pt}}$}}

\def\smallheads{
	\renewcommand{\Large}{\large} 
	\renewcommand{\LARGE}{\Large}} 
\standard

\smallheads
\newtheorem{procedure}{Procedure}
\newtheorem{definition}{Definition}
\newtheorem{proposition}{Proposition}

\newtheorem{corollary}{Corollary}

\newtheorem{example}{Example}

\newtheorem{remark}{Remark}

\def\5n{\negthinspace \negthinspace \negthinspace \negthinspace \negthinspace }
\def\4n{\negthinspace \negthinspace \negthinspace \negthinspace }
\def\3n{\negthinspace \negthinspace \negthinspace }
\def\2n{\negthinspace \negthinspace }

\def\0{\mathbf{0}}
\def\1{\mathbf{1}}
\def\E{\mathbb{E}}

\def\x{\bm x}

\begin{document}
\graphicspath{{figures/}}

\begin{titlepage}
	
	\centerline{\Large\bf Collective Wisdom: Policy Averaging,} 
	\medskip
	\centerline{\Large\bf with an Application to the Newsvendor Problem}
	
	\medskip
	
	\centerline{\Large\bf }
	
	\vspace{2in}
	
	\centerline{Xiangyu Cui*}
	
	\vspace{.15in}
	
	\centerline{Nicholas G. Hall$^{\dagger}$}
	
	\vspace{.15in}
	
	\centerline{Yun Shi$^{\ddagger}$}
	
	\vspace{.15in}
	
	\centerline{Tianyuan Su*}
	
	\vspace{1.4in}
	
	\noindent * School of Statistics and Data Science, Shanghai University of Finance and Economics, Shanghai, China
	
	\vspace{.15in}
	
	\noindent $^{\dagger}$ Department of Operations \& Business Analytics, Fisher College of Business, The Ohio State University, Columbus, Ohio, USA
	
	\vspace{.15in}
	
	\noindent  $^{\ddagger}$ School of Statistics, Academy of Statistics and Interdisciplinary Sciences,  East China Normal University, Shanghai, China
	
	\vspace{1.3in}
	
	\centerline{November 22, 2024}
	
\end{titlepage}

\pagestyle{empty}%
\baselineskip=21pt

\vspace{.3in}

\centerline{\Large\bf Abstract}

\vspace{.2in}

We propose a Policy Averaging Approach (PAA) that synthesizes the strengths of existing approaches to create more reliable, flexible and justifiable policies for stochastic optimization problems. An important component of the PAA is risk diversification to reduce the randomness of policies. A second component emulates model averaging from statistics. A third component involves using cross-validation to diversify and optimize weights among candidate policies. We demonstrate the use of the PAA for the newsvendor problem. For that problem, model-based approaches typically use specific and potentially unreliable assumptions of either independently and identically distributed (i.i.d.) demand or feature-dependent demand with covariates or autoregressive functions. Data-driven approaches, including sample averaging and the use of functions of covariates to set order quantities, typically suffer from overfitting and provide limited insights to justify recommended policies. By integrating concepts from statistics and finance, the PAA avoids these problems. We show using theoretical analysis, a simulation study, and an empirical study, that the PAA outperforms all those earlier approaches. The demonstrated benefits of the PAA include reduced expected cost, more stable performance, and improved insights to justify recommendations. Extensions to consider tail risk and the use of stratified sampling are discussed. Beyond the newsvendor problem, the PAA is applicable to a wide variety of decision-making problems under uncertainty. 

\vspace{.35in}


\vspace{.2in}

\noindent {\bf 1989 OR/MS Subject Classification:} \\
Operations management; \\
Stochastic decision making.

\newpage

\pagestyle{plain}%
\pagenumbering{arabic}

\section{Introduction to Policy Averaging}  \label{sec:PAAintro} 

\begin{figure}[htbp]
\centering
\includegraphics[width=0.75\textwidth]{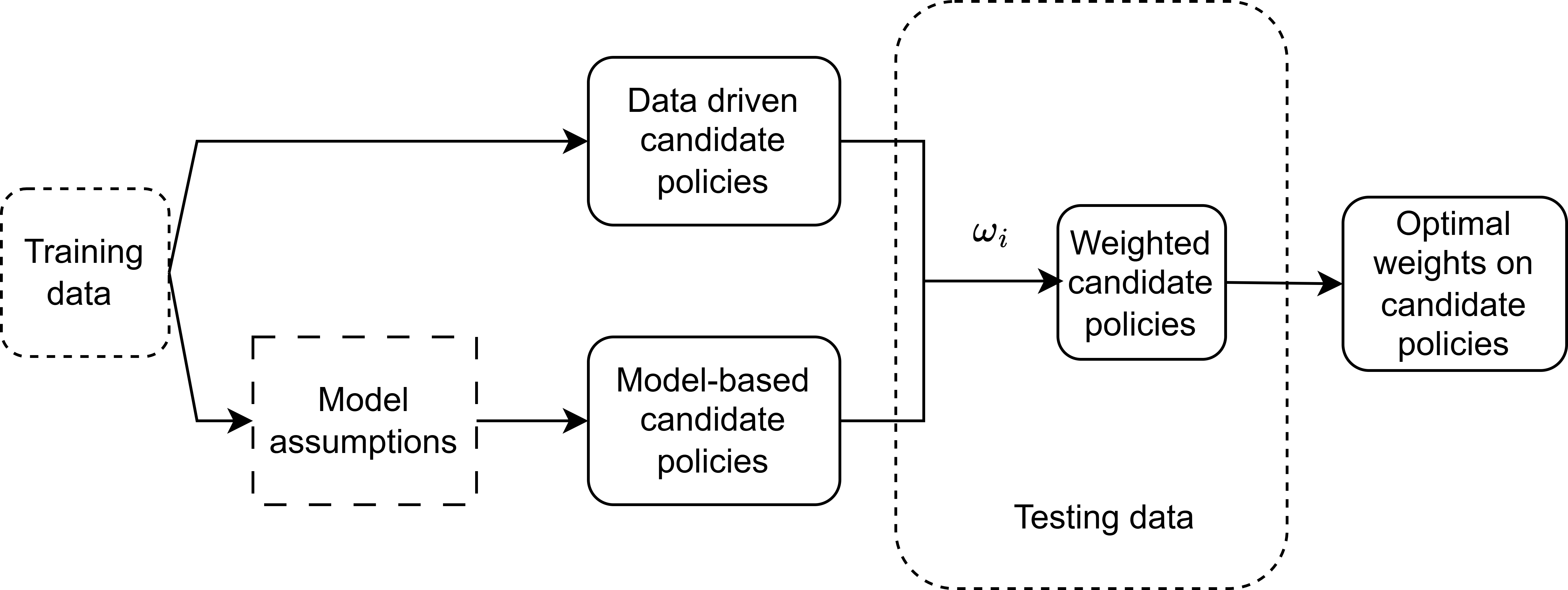} 
\caption{Overview of the Policy Averaging Approach}
\label{fig:paa}
\end{figure}

Existing approaches for solving stochastic optimization problems can be classified into two categories. First,  {\em model-based approaches} use specific and potentially unreliable assumptions of either i.i.d. data or feature-dependent data based on covariates or autoregressive functions. Second, utilizing the availability of large data sets, alternative {\em data-driven approaches} have recently gained prominence. Data-driven approaches, including sample averaging and the use of functions of covariates to recommend decisions, base their policies on historical data. However, overfitting is common, and limited insights are provided to justify recommended policies. Because of these widely observed shortcomings of both existing approaches, we propose an alternative, the Policy Averaging Approach (PAA). The overall motivation for our work is to improve decision making under uncertainty. As we explain below, the PAA is potentially applicable to a wide variety of optimization problems under uncertainty. An overview of the PAA is illustrated in Figure \ref{fig:paa}, with further details provided in the following paragraphs.

We first describe the PAA as a methodology for stochastic optimization in general, without reference to any specific application, starting with several definitions. An {\em approach} is a general methodology, for example a {\em model} or a {\em collection of models} for solving an optimization problem. It takes as input a set of {\em data} and provides as output a {\em policy} which is a numerical recommendation for one or more decisions for the optimization problem. 

The PAA starts by generating $m$ different numerical {\em candidate policies}. These policies may be generated from the use of different models, or the use of different data sets, or both. For example, if the stochastic optimization problem is a simple one with a closed form solution, then only one model is typically relevant; however,  different data sets may still generate different policies that we may diversify. At the other extreme, if the optimization problem is intractable, alternative heuristic solution approaches can be used as models, and also applied to different data sets, potentially resulting in a larger set of candidate policies.

Each candidate policy is treated as a random variable or a random function, rather than as deterministic. In principle, the PAA adjusts the weighting over those policies to optimize the (typically, expected cost or profit) objective of the problem. These weights provide diversification over the set of candidate policies. However, since both (a) the true distribution of the data, and (b) the true joint distribution of the candidate policies, are unknown, the problem of optimization over the weights cannot be solved directly.

Consequently, we derive the {\em empirical counterparts} of the distribution of the data and of the joint distribution of the policies, from the given data. To accomplish this, we repeatedly use the data with one part omitted to estimate $m$ candidate policies and minimize a total weighted policy function for the omitted data, similar to the mechanism of Leave-One-Out Cross-Validation (\citealp{Witten2011data}). We use a ``training'' portion of the data to identify the relevant models where necessary and then to compute the numerical candidate policies. The remaining ``testing'' data is then used to evaluate the candidate policies and determine the best weights for them. One result of this procedure is that every candidate policy is a special case of the policy recommended by the PAA. This approach has potential application to a wide variety of stochastic optimization problems.

The PAA combines concepts from both finance and statistics. A key component is risk diversification (\citealp{wagner1971effect}) from finance, which involves holding a variety of securities to reduce the impact of uncertainty in their outcomes. In the context of the PAA, this principle is applied to diversify the uncertainty associated with candidate policies. A second key component comes from model averaging (\citealp{draper1995assessment}) in statistics, which combines results from a family of models to reduce forecasting errors. Typically, model averaging requires the models to have similar structures — such as linear models — and the weights to be positive and less than 1. However, our goal here is not to achieve more accurate estimation, as in traditional model averaging, but to focus on more effective optimization. Consequently, we allow more general structural properties and weights. This enhances the flexibility of our approach, as demonstrated by our theoretical, simulation, and empirical results. A third component of the PAA is the use of ``testing'' data to optimize the weights, thereby diversifying among policies, similar to the cross-validation concept from statistics and machine learning (\citealp{Witten2011data}).

The contributions of this work are as follows. For general application to optimization problems under uncertainty, we describe a Policy Averaging Approach that is based on synthesizing the strengths, and mitigating the weaknesses, of  existing approaches in the literature. For the classical newsvendor problem in particular, we prove that the PAA recommends decisions that are at least as good as those of the individual approaches proposed earlier. An extensive simulation study shows that the performance of the PAA is better than that of any earlier model, with respect to minimizing expected cost. The  observed improvement in performance is strongest in cases of misspecification of random data. We also demonstrate strong outperformance of the PAA on an empirical (real world) newsvendor data set with feature-dependent data. We also discuss the mitigation of tail risk and the use of stratified sampling over different categories of solution approach for generating candidate policies.

Section~\ref{sec:newsvendorintro} describes traditional model-based and data-driven approaches to the newsvendor problem, and provides several detailed examples. Section~\ref{sec:PAANewsvendor} contains two parts. The first part applies the PAA to the newsvendor problem under i.i.d. demand, by optimization of weights over various demand quantile estimates. The second part extends the PAA to the more general case of feature-dependent demand, which requires optimization of weights over functions of covariates. Section~\ref{sec:simulation} conducts extensive comparisons by simulation that show the substantial improvement of the PAA over existing approaches. In Section~\ref{sec:empirical}, we demonstrate similar improvement by the PAA for a real-world data set for the newsvendor problem. Section~\ref{sec:extensions} discusses the control of tail risk in the PAA and the use of stratified sampling. Section~\ref{sec:conclusions} provides a summary and some suggestions for future work. Proofs of all the theoretical results appear in an Appendix.

\section{Policies for the Newsvendor Problem}  \label{sec:newsvendorintro}

The newsvendor problem, a foundational problem in stochastic inventory theory, provides a framework for understanding inventory management decisions, and more generally stochastic optimization. This problem requires determining an optimal order quantity to minimize the total costs associated with overordering and underordering, given a single period with demand uncertainty. The basic newsvendor model has been extended in various directions, including to multiple periods with various assumptions about lost sales, to enhance its range and applicability (\citealp{khouja1999single}). 
The newsvendor literature includes two main approaches to address this classic  problem. One approach is model-based approach and the other approach is data-driven approach. These approaches are  discussed in Sections~\ref{subsec:newsvendormodels} and \ref{subsec:newsvendordd}, respectively.

\subsection{Model-based approaches}  \label{subsec:newsvendormodels}

A model-based approach for newsvendor problem relies on explicitly specifying a probabilistic distribution to characterize the demand. Model-based approaches to the newsvendor problem either assume an i.i.d. demand or take into account the influence of features such as covariates on demand. Under the i.i.d. assumption, the normal distribution is a prevalent choice due to its flexibility and ease of sensitivity analysis (\citealp{silver1998inventory}). Alternative distributions have also been studied in this context. \cite{khouja1996two} use an exponential distribution to illustrate emergency supply effects.  \cite{huang2011competitive} explore various distributions including normal, Poisson and exponential distributions to address multi-product inventory problems with dynamic availability.  \cite{levi2015data} focus on using log-concave distributions, including common forms,  such as normal, uniform, exponential, logistic, and chi-square, distributions to study the newsvendor problem.

Beyond the i.i.d.~assumption, the literature has also explored feature-based models, where demand depends on covariates or exhibits correlations over time. Here, a covariate \( \x_t \) can represent various factors influencing demand. For example, \( d_{t-1} \), the demand at time \( t-1 \), can model the time-correlated structure of demand  (\citealp{aviv2002gaining}). \cite{johnson1975optimality} introduce ARIMA models to evaluate the optimality of myopic inventory policies under dependent demand. A study by \cite{lee2000value} extends AR(1) models to address complexities in demand, such as trends, seasonality, and non-stationarity. The Martingale Model of Forecast Evolution (MMFE), developed independently by \cite{heath1994modeling} and \cite{graves1998dynamic}, has become a foundational framework for modeling dynamic forecast updates. This model was later applied to inventory systems by \cite{aviv2001effect}, providing valuable insights into inventory management under evolving demand conditions.

Beyond traditional time-series models, the incorporation of covariates significantly advances demand modeling. \cite{olivares2008structural} apply structural estimation to account for demand heterogeneity through explanatory variables. \cite{beutel2012safety} introduce regression-based models that incorporate exogenous factors such as price and weather.  \cite{green2013nursevendor} include workload-dependent absenteeism as a critical factor in nurse staffing models. 

We now provide the basic mathematical formulation and notations for the obtained model-based policies for both the i.i.d. and feature-based cases.  For a model-based approach, the demand \( d \) is assumed to follow a probability distribution function (PDF) \( f(d; \x, \bm \theta) \), where \( \x \) is the vector of covariates, and $\bm \theta$ is a $p$ dimensional vector of the distribution parameters.  Then, based on the data set $\bm D_t = \{(d_1,\bm x_1),(d_2,\bm x_2),\ldots,(d_t,\bm x_{t})\}$, the decision maker faces the following optimization problem, 
 \begin{align}\label{eq:MD}
\min_{Q}~ c_o \int_{-\infty}^{Q}(Q-y) f(y;\x, \bm \theta) d y+c_u \int_{Q}^{+\infty}(y-Q) f(y;\x, \bm \theta) dy, 
 \end{align}
where \( c_o \) and \( c_u \) represent the known costs of overage and underage, respectively, $d_i$ and $\bm x_{i}$ denote the observed demand and co-variates for $i$th period, respectively, and the objective is the conditional expected cost. When the covariate \(\bm x_{i}\) is a deterministic constant, the setting is the i.i.d. case. Otherwise, it is the feature-based case.

To solve the model-based optimization problem (\ref{eq:MD}) and obtain the model-based policy, we can use two different types of computational approaches, the estimate-then-optimization (ETO) approach and the integrated-estimation-optimization (IEO) approach (\citealp{elmachtoubgMS2022}, \citealp{elmachtoub2023estimate}). Under the ETO approach, the decision maker first obtains the estimated parameter \(\hat{\bm \theta}(\bm D_t)\) based on the data by statistical analysis, such as maximum likelihood method, and then solves the optimization problem for a value of the covariate $\bm x$,
\begin{align}\label{eq:ETO}
(ETO)~~\min_{Q}~ c_o \int_{-\infty}^{Q}(Q-y) f(y;\x, \hat{\bm \theta} (\bm D_t)) d y+c_u \int_{Q}^{+\infty}(y-Q) f(y;\x, \hat{\bm \theta} (\bm D_t)) dy.
 \end{align}
We denote the derived optimal policy as $Q^{ETO}(\bm x; \hat{\bm \theta} (\bm D_t))$. When changing the values of covariate $\bm x$, $Q^{ETO}(\bm x; \hat{\bm \theta} (\bm D_t))$ characterizes a deterministic function of $\bm x$.  

Under the IEO approach, the decision maker  first solves the optimization problem (\ref{eq:MD}) and obtains the dependence structure of order quantity with respect to covariates and parameters, which is denoted as $Q^{IEO}(\bm x, \bm \theta)$, where $Q^{IEO}(\cdot,\cdot)$ is a deterministic function of both $\bm x$ and $\bm \theta$. Then, the decision maker determines the best parameter, $\tilde{\bm \theta}(\bm D_t)$,  following the optimization problem,\begin{align}\label{eq:IEO}
 (IEO)~~\min_{\bm \theta}~  \frac{c_o}{t}\sum_{j=1}^t \max (Q^{IEO}(\bm x_{j}, \bm\theta)-d_j,0) + \frac{c_u}{t}\sum_{j=1}^t \max (d_j-Q^{IEO}(\bm x_{j}, \bm\theta),0),
 \end{align}
whose objective is the empirical cost over the data. 
Finally, we have the optimal policy as
\(Q^{IEO} (\bm x, \tilde{\bm \theta} (\bm D_t))\) for the covariate $\bm x$.

We divide the model-based approaches into two categories from another perspective: well-specified and misspecified approaches. Although the terms ``well-specified'' and ``misspecified'' are defined in the context of the model used in these approaches, we also refer to the corresponding approaches as well-specified or misspecified when there is no ambiguity. Their precise definitions are as follows. 

\begin{definition}[Well-Specified]
\label{def:well_specified}
A model-based approach is well-specified, if there exists a $\bm \theta \in \mathbb{R}^{p}$ such that $f(d;\bm x, \bm \theta) = g(d;\bm x)$, where $g(d;\bm x)$ is the real unknown distribution of demand. 
\end{definition}

\begin{definition}[Misspecified]
\label{def:mis_specified}
A model-based approach is misspecified, if for any $\bm \theta \in \mathbb{R}^{p}$, $f(d;\bm x, \bm \theta) \neq g(d;\bm x)$ holds for some $\bm x$, where $g(d;\bm x)$ is the real unkown distribution of demand. 
\end{definition}

The main advantage of the obtained model-based policies is their ability to provide interpretable order quantities, efficiently utilizing data to yield accurate results when the approach is well-specified. However, if the approach is misspecified, these policies often lead to inaccurate order quantities. Moreover, it is typically challenging to determine in advance whether the approach is well-specified or misspecified in practice.

\subsection{Data-driven approaches}  \label{subsec:newsvendordd}
The second data-driven approach develops data-driven policies, and requires no explicit assumption about the demand distribution, but instead relies solely on the observed demand data to provide information for decision-making. The simplest approach of this type is the sample averaging approach (SAA), which assumes that the empirical distribution of demand data is the real demand distribution and determines a constant order quantity policy. SAA implicitly assumes that the demand in different periods are i.i.d. but without an explicit distributional assumption. 

In the presence of covariates $\bm x_t$, data-driven approaches aim to find the optimal order quantity as a function of $\bm x_t$ directly. \cite{huber2019data} sets the order quantity as a function of the covariates $\bm x_t$ and finds the coefficients by quantile regression. \cite{ban2019big} set the order quantity as a polynomial function of the covariates, where the coefficients are determined by empirical risk minimization (ERM). \cite{oroojlooyjadid2020applying} and \cite{cao2019quantile} set the order quantity as a general function of the covariate, as modeled by an artificial neural network (ANN) and a double parallel feedforward network (DPFNN), respectively, where the coefficients are determined by optimizing the performance on the data set.  \cite{ban2019big} also propose the order quantity as an implicit function of the covariates, which is obtained by ERM with kernel weights (KO) following the formula for Nadaraya-Watson kernel regression (\citealp{nadaraya1964estimating}, \citealp{watson1964smooth}).

We also provide the basic mathematical formulation and notations for data-driven approaches.  Based on the data set $\bm D_t = \{(d_1,\bm x_1),(d_2,\bm x_2),\ldots,(d_t,\bm x_{t})\}$,  the decision maker faces the following optimization problem,
\begin{align}\label{eq:DD} \min_{\bm q }~ & \frac{c_o}{t}\sum_{j=1}^t \max (h(\bm x_{j};\bm q)-d_j,0) +\frac{c_u}{t}\sum_{j=1}^t \max (d_j-h(\bm x_{j};\bm q),0)+\lambda \|\bm q\|_{k}^2,
\end{align}
where $h(\cdot~;\bm q)$ is a deterministic function, $\bm q$ is the parameter of the function, $\|\bm q\|_k^2$ denotes the $\ell_k$ norm of the vector $\bm q$, which is the regularization term, and $\lambda\geq 0$ represents the weights on the regularization term. 

When $h(\bm x_{j}; \bm q)$ is a constant function and $\lambda = 0$, the optimization problem (\ref{eq:DD}) determines the optimal policy of the SAA approach, $Q^{SAA}(\bm D_t)$. When $\lambda=0$ and the order quantity is set as a polynomial function of the covariates, such as $h(x_{j};\bm q) = \sum_{i=0}^{p} q_i x_{j}^i$ for a one-dimensional covariate, where $p$ is the highest order of the function and $\bm q = (q_0,q_1,\cdots,q_p)^T$ is the coefficient vector of the polynomial function, the optimization problem (\ref{eq:DD}) gives the optimal policy of the ERM approach, $Q^{ERM}( x; \tilde{\bm q}(\bm D_t)) = h(x; \tilde{\bm q}(\bm D_t))$, where $\tilde{\bm q}(\bm D_t)$ is the optimizer of problem (\ref{eq:DD}). When $\lambda>0$ and the order quantity is set as a polynomial function of the covariates, optimization problem (\ref{eq:DD}) gives the optimal policy of the ERM approach with regularization, $Q^{ERM-\ell_k}(x; \tilde{q}(\bm D_t))$. When the order quantity is set to the following function,
\begin{align*}
h(\bm x_{j};\bm q) = \varphi_{n+1} ( \bm w^{(n+1)}\bm \varphi_{n} (\bm w^{(n)} \dots \bm \varphi_{2} (\bm w^{(2)} \bm \varphi_{1} (\bm w^{(1)} \bm x_{j} + \bm b^{1}) + \bm b^{(2)})\dots + \bm b^{n}) + \bm b^{(n+1)}),
\end{align*}
where $n$ denotes the depth of the neural network, $\bm\varphi_{j}(\cdot)$ is the activate function with appropriate dimension of layer $j$, $\bm w^{(j)}$ and $\bm b^{(j)}$ are the parameters for layer $j$, $\varphi_{n+1}$ is the activate function of the output layer, and $\bm q$ contains all the parameters representing the neural network. Then, the optimization problem (\ref{eq:DD}) gives the policies of ANN, DNN and DPFNN with additional structures on the network, $Q^{ANN}(\bm x; \tilde{\bm q}(\bm D_t))$, $Q^{DNN}(\bm x; \tilde{\bm q}(\bm D_t))$ and $Q^{DPFNN}(\bm x; \tilde{\bm q}(\bm D_t))$. 
 
When we change the problem (\ref{eq:DD})  by adding kernel-weights for each demand sample as follows
\begin{align}\label{eq:DD-2}(KO)~~\min_{Q}~ & \frac{c_o}{t}\sum_{j=1}^t \frac{K_b\left(\bm x-\bm x_j\right)\max (Q -d_j,0)}{\sum_{i=1}^t K_b\left(\bm{x}-\bm{x}_i\right)}  +\frac{c_u}{t}\sum_{j=1}^t \frac{K_b\left(\bm x-\bm x_j\right)\max (d_j-Q,0)}{\sum_{i=1}^t K_b\left(\bm{x}-\bm{x}_i\right)},
\end{align}
where $K_b(\cdot)$ is the kernel function with bandwidth $b$, the optimizer of problem (\ref{eq:DD-2}) is a scalar, which depends on the covariate $\bm x$ and the data set $\bm D_t$, which construct an implicit mapping from $\bm x$ to an optimal order quantity. Then,  we obtain  the policy as $Q^{KO}(\bm x;\bm D_t)$. 

Data-driven approaches easily incorporate the complex structure between the demand and the covariates and efficiently use the data. However, there are also weaknesses of data-driven policies, include overfitting, and limited insights from the results which makes them harder for decision makers to justify in practice.

 \section{Policy Averaging Approach for the Newsvendor Problem}
\label{sec:PAANewsvendor}

For a finite data set, the order quantity policies of both model-based approaches and data-driven approaches are random. We adopt the philosophy of risk diversification in financial portfolio theory, and propose a policy averaging approach (PAA) for the newsvendor problem. The PAA considers the order quantity policies of different approaches in the literature as securities and finds a ``portfolio'' of these policies, i.e., the weighted summation of existing policies, to obtain better performance. 

Formally, the PAA problem is:
\begin{align*}
(PAA) \quad \min_{\{\omega_1, \ldots, \omega_m\}} \quad & \mathbb{E}[C(Q - d)] \\
\text{s.t.} \quad & Q = \sum_{i=1}^m \omega_i Q^{(i)}(\bm D_t), \\
& L \leq \omega_i \leq U, \quad i = 1, \ldots, m, \\
& \sum_{i=1}^m \omega_i = 1,
\end{align*}
where \( C(Q - d) = c_o \max(Q - d, 0) + c_u \max(d - Q, 0) \) is the cost function associated with an order quantity \( Q \) and demand \( d \), \( Q^{(1)}(\bm D_t), \ldots, Q^{(m)}(\bm D_t) \) are random candidate policies based on data $\bm D_t$, and the expectation operator \( \mathbb{E}[\cdot] \) is applied to both the random candidate policies and the random demand \( d \). The weights \( \omega_i \) are constrained by \( L \leq \omega_i \leq U \), where \( L \leq 0 \) and \( U \geq 1 \) represent the lower and upper bounds, respectively. We allow \( L \) to take negative values and \( U \) to exceed 1. This strategy is validated by our simulation results in Section~\ref{sec:simulation}. 
We denote the PAA policy as $Q^{PAA} = \sum_{i=1}^{m} \omega_i^* Q^{(i)}(\bm D_t)$, where $\omega_i^*$ is the optimizer of problem $(PAA)$.

The random policies under the i.i.d. case are random variables, while they are random functions under the feature-based case. We provide detailed discussions for the two cases, respectively. Section~\ref{subsec:PAAiid} applies the PAA to newsvendor problems with i.i.d. demand. Section~\ref{subsec:PAAcorrdem} applies the PAA for the newsvendor problem with feature-based demand. 


 \subsection{The i.i.d. demand case} \label{subsec:PAAiid}
In this section, we assume that the observed demand data are i.i.d., with an unknown cumulative distribution function (CDF) denoted by \(G(d)\), and there are \(m\) candidate policies, \(Q^{(1)}(\bm D_t), \ldots, Q^{(m)}(\bm D_t)\), which may be generated from either model-based approaches or data-driven approaches. As the candidate policies are either computed directly from finite data, such as $Q^{SAA}(\bm D_t)$, or a function of parameters estimated from finite data such as \( Q^{ETO}(\hat{\bm \theta}(\bm D_t)) \) or \( Q^{IEO}(\tilde{\bm \theta}(\bm D_t))\), they are all random variables.  We denote the theoretical optimal order quantity as $Q^{T-opt} = G^{-1}\big(\frac{c_u}{c_o+c_u}\big)$, which is an unkown constant.  We omit the parameters $\hat{\bm \theta}(\bm D_t)$,  $\tilde{\bm \theta}(\bm D_t)$ notations in the candidate policies to simplify the formulations. We further assume that there is a joint CDF \(F(Q^{(1)}(\bm D_t), \ldots, Q^{(m)}(\bm D_t))\) for these random candidate policies, which is not known by the decision maker.

As $G(d)$ and \(F(Q^{(1)}(\bm D_t), \ldots, Q^{(m)}(\bm D_t))\) are unknown to the decision maker, problem $(PAA)$ cannot be solved directly. Hence, we propose the following procedure to solve $(PAA)$.

\begin{procedure}\label{proc:iid}{\bf [Policy averaging approach for i.i.d. case]}\\
1. Define the reduced data set without $d_j$ as $\bm D_{\cancel{j}}=\{d_1,\ldots,d_{j-1},d_{j+1},\ldots,d_t\}$.\\
2. Compute the $m$ candidate policies over the reduced data set $\bm D_{\cancel{j}}$, $Q^{(1)}(\bm D_{\cancel{j}})$, $\ldots$, $Q^{(m)}(\bm D_{\cancel{j}})$.\\
3. Determine the optimal weights $\omega_1^*,\cdots,\omega_m^*$by minimizing the average cost over unused data, \begin{align*}
\min_{\{\omega_1,\cdots,\omega_m\}}~\textstyle\sum_{j=1}^t C(\sum_{i=1}^{m}\omega_i Q^{(i)}(\bm D_{\cancel{j}})-d_j).
\end{align*}
\end{procedure}

By adding the constraints for weights, we have the following optimization problem $(PAA_{iid})$, 
\begin{align*}
 (PAA_{iid}) \quad \min_{\{\omega_1,\cdots,\omega_m\}} \quad & \frac{c_o}{t} \sum_{j=1}^t \max (Q(\bm D_{\cancel{j}}) - d_j, 0) + \frac{c_u}{t} \sum_{j=1}^t \max (d_j - Q(\bm D_{\cancel{j}}), 0) \\
 \text{s.t.} \quad & Q(\bm D_{\cancel{j}}) = \sum_{i=1}^m \omega_i Q^{(i)}(\bm D_{\cancel{j}}), \\
 & L \leq \omega_i \leq U, ~i=1,\ldots,m, \\
 & \sum_{i=1}^m \omega_i = 1, \end{align*}
where $L\leq 0$ and $U\geq 1$. The optimization problem $(PAA_{iid})$ is convex. We denote the PAA policy as $Q^{PAA}(\bm D_t) =\sum_{i=1}^{m}\omega_i^* Q^{(i)}(\bm D_t)$. Furthermore, the PAA approach integrates the cross-validation idea into the optimization procedure. We use a portion of the data to compute the candidate policies, and then decide the optimal weights based on the unused testing data. If $\omega_i \in \{0,1\}$, the optimization problem $(PAA_{iid})$ is reduced to a horse race among the candidate policies on the testing data. Since the demand \(d_j\) is independent of the data in \(\bm{D}_{\cancel{j}}\), the candidate order quantities \(Q^{(i)}(\bm{D}_{\cancel{j}})\) and \(d_j\) are independent. This implies that the demand \(d_j\) and the order quantity \(Q(\bm{D}_{\cancel{j}})\) in problem \((PAA_{iid})\) are independent, providing a good approximation of \((PAA)\) on the given dataset \(\bm{D}_t\). In Procedure \ref{proc:iid}, we may also use a batch size other than 1 for the testing set. The discussion of batch size selection is left as future work.

After proposing the PAA policy, we investigate its theoretical properties. The asymptotic properties of PAA policy are given in Proposition~\ref{iidprop:PAA is the best} and Proposition~\ref{iidprop:PAA converges to the optimal}. The finite sample properties of PAA policy are given in Proposition~\ref{iidprop:corr dis}, Proposition~\ref{iidprop:mean dis} and Proposition~\ref{iidprop:impro from dis}.

\begin{proposition}  
\label{iidprop:PAA is the best}
When the demand data are independent and identically distributed (i.i.d.) and the sample size tends to infinity ($t\to \infty$), the PAA policy achieves an expected cost that is no greater than the expected cost of any candidate policy.
\end{proposition}
 
\begin{proposition}
\label{iidprop:PAA converges to the optimal}
Under the following assumptions:
\begin{itemize}[itemsep=-1pt]
\item The demand data are independent and identically distributed (i.i.d.);
\item The candidate policy \(Q^{(i)}(\bm D_t)\) converges to a constant \(\bar{Q}^{(i)}\) as the sample size \(t \to \infty\), with a standard deviation of \(\frac{M_i}{\sqrt{t}}\) for \(i = 1, \dots, m\);
\item \(\bar{Q}^{(i)} \neq \bar{Q}^{(j)}\) for some \(i \neq j\);
\item The lower bound \(L\) and upper bound \(U\) for data set $\bm D_t$ approach negative and positive infinity, respectively, at a rate slower than \(-\sqrt{t}\) and \(\sqrt{t}\);
\end{itemize}
then the PAA policy $Q^{PAA}(\bm D_t)$ converges to the true optimal order quantity \(Q^{T-opt}\) as the sample size $t\to \infty$.
\end{proposition}

\begin{corollary} 
\label{iidcor:PAA converge even all mis}
Under the four assumptions in Proposition \ref{iidprop:PAA converges to the optimal}, even if all the models used in the candidate policies are misspecified, and \( \bar{Q}^{(i)} \neq Q^{T-opt}\) for all \( i \), the PAA policy still converges to the true optimal order quantity \( Q^{T-opt}\). This is because the expanded search space for the order quantity, facilitated by the increasing bounds, guarantees that the true optimal order quantity is included within the search region.
\end{corollary}

The asymptotic properties of the PAA policy demonstrate its effectiveness as the sample size increases. The assumption in Proposition \ref{iidprop:PAA converges to the optimal} that candidate policies converge to constant values as the sample size tends to infinity is typically satisfied, as we now discuss. In model-based approaches, candidate policies are functions of model parameters, which converge to their true values as the sample size increases, provided the model is well-specified. \cite{white1982maximum} shows that parameters estimated via maximum likelihood estimation consistently converge to the true constant parameters of a specified model, minimizing the Kullback–Leibler divergence between the specified and true models, even in the case of model misspecification. In data-driven approaches such as SAA, candidate policies also converge to the true optimal order quantity under the i.i.d. assumption. Additionally, the technical condition on the lower bound 
$L$ and upper bound 
$U$ ensures that the variance of the combined PAA policy approaches zero as the sample size grows, guaranteeing a stable estimation of the order quantity at the asymptotic limit. 
 
In addition to the asymptotic properties of the PAA policy, we now examine its finite-sample properties with the sample size \( t \) fixed. In the following propositions, we consider only the combination of two candidate policies, \( Q^{(1)}(\bm{D}_t) \) and \( Q^{(2)}(\bm{D}_t) \), which are random variables. However, the general case with multiple candidate policies can easily be extended by incrementally adding each candidate policy to the resulting weighted policy. 
For illustration, we provide the expression for the expected cost of a candidate policy \( Q^{(i)}(\bm{D}_t) \), \( i = 1, 2 \), which follows a normal distribution \( N(\mu_i, \sigma_i^2) \), under the assumption that the demand is also normally distributed as \( N(\mu_d, \sigma_d^2) \) and the demand and candidate policies are independent. The expected cost is given by:
\begin{align*}
  \mathbb{E}[C(Q^{(i)}(\bm{D}_t) - d)] = ~& (c_o + c_u) \frac{\sqrt{\sigma_i^2 + \sigma_d^2}}{\sqrt{2\pi}} e^{-\frac{(\mu_i - \mu_d)^2}{2(\sigma_i^2 + \sigma_d^2)}} + (\mu_i - \mu_d) (c_o - c_u) \\
  & + (\mu_i - \mu_d) \Big[ c_u \Phi\Big(\frac{\mu_i - \mu_d}{\sqrt{\sigma_i^2 + \sigma_d^2}} \Big) - c_o \Phi\Big( \frac{\mu_d - \mu_i}{\sqrt{\sigma_i^2 + \sigma_d^2}} \Big) \Big],
\end{align*}
where \( \Phi(\cdot) \) denotes the cumulative distribution function of the standard normal distribution. We denote this expected cost as \( J(\mu_i, \sigma_i^2) \), which is a function of \( \mu_i \) and \( \sigma_i^2 \). The detailed computation of the expected cost can be found in the proof of Proposition \ref{iidprop:corr dis}.

\begin{proposition}
\label{iidprop:corr dis}
For a fixed sample size \( t \), consider two candidate policies, \( Q^{(1)}(\bm{D}_t) \) and \( Q^{(2)}(\bm{D}_t) \), that follow a two-dimensional normal distribution 
\(N\left( \begin{pmatrix} \mu \\ \mu \end{pmatrix}, \begin{pmatrix} \sigma^2 & \rho \sigma^2 \\ \rho \sigma^2 & \sigma^2 \end{pmatrix} \right)
\), where the correlation coefficient \( \rho < 1 \). The demand follows a normal distribution \( N(\mu_d, \sigma_d^2) \), and the demand and candidate policies are assumed to be independent.

Under these conditions, the optimal weights for the candidate policies are \( \omega_1^* = \omega_2^* = \frac{1}{2} \). Also, the improvement in the expected cost of the PAA policy, \( Q^{\text{PAA}}(\bm{D}_t) \), over the individual candidate policies is given by \( J(\mu, \sigma^2) - J(\mu, (1+\rho)\sigma^2/2) \).
\end{proposition}

Under the assumptions of Proposition~\ref{iidprop:corr dis}, the PAA policy has the same expected value as the candidate policies, but it exhibits a smaller variance, leading to a lower expected cost. This phenomenon is consistent with the concept of risk diversification among imperfectly correlated risky assets in financial portfolio theory. In this analysis, the correlation coefficient \( \rho < 1 \) plays a crucial role. As \( \rho \) decreases, the improvement in the expected cost of the PAA policy increases. Although the demand distribution parameters \(\mu_d\) and \(\sigma_d\) do not play an explicit role in Proposition \ref{iidprop:corr dis}, they implicitly influence the distribution parameters of candidate policies, \(\mu\), \(\sigma\), and \(\rho\), due to the methods used to obtain those candidate policies.
 
\begin{figure}[htbp]
	\centering
		\begin{subfigure}[b]{0.3\textwidth}
			\centering
			\includegraphics[width=\textwidth, height=0.2\textheight]{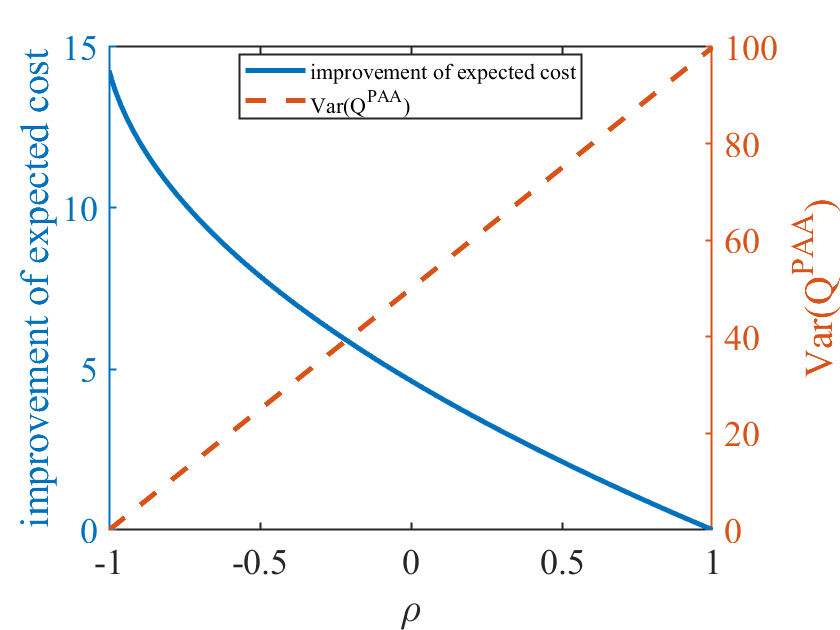}
			\caption{$\sigma_d^2=1$}
			\label{fig:corr dis:subfig:a}
		\end{subfigure}
		\hfill
		\begin{subfigure}[b]{0.3\textwidth}
			\centering
			\includegraphics[width=\textwidth, height=0.2\textheight]{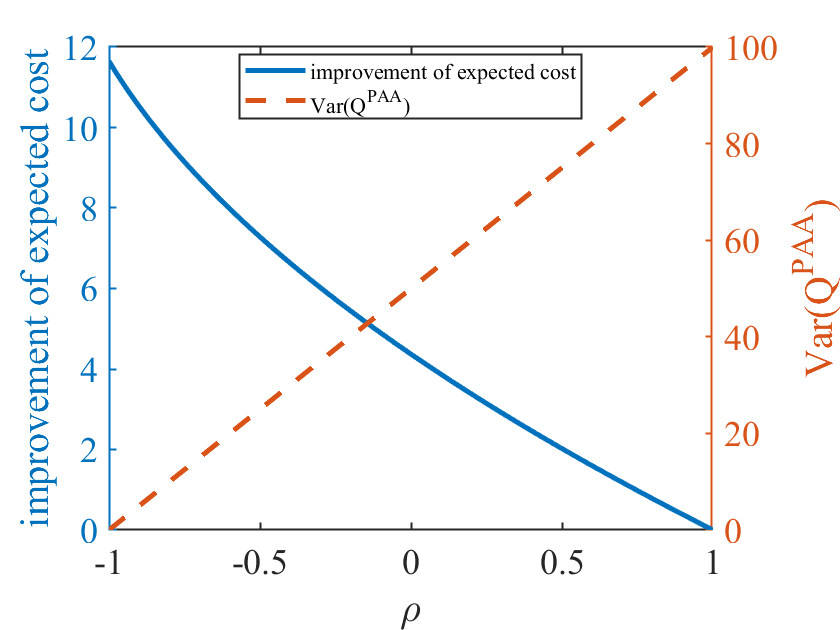} 
			\caption{$\sigma_d^2=10$}
			\label{fig:corr dis:subfig:b}
			
		\end{subfigure}
		\hfill
            \begin{subfigure}[b]{0.3\textwidth}
			\centering
			\includegraphics[width=\textwidth, height=0.2\textheight]{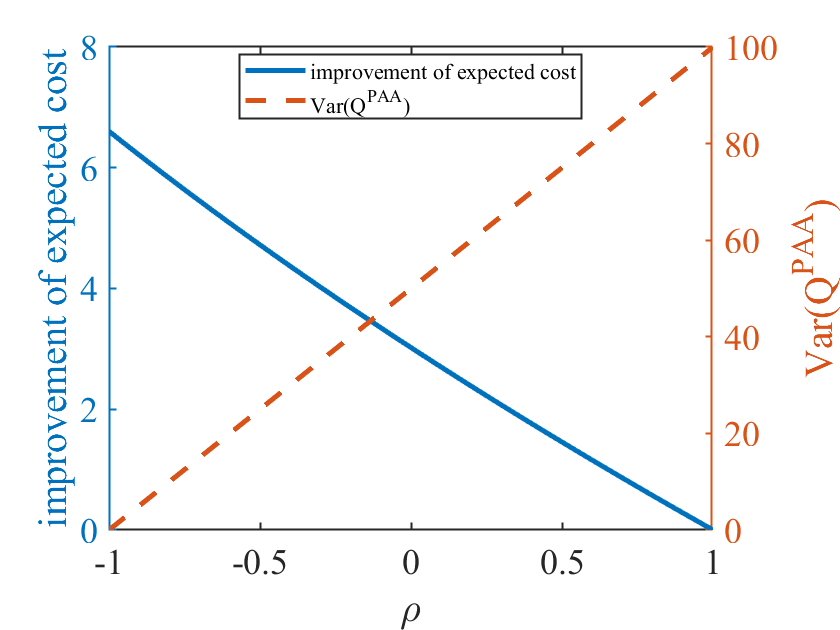} 
			\caption{$\sigma_d^2=100$}
			\label{fig:corr dis:subfig:c}
		\end{subfigure}
		\hfill
		\caption{Effect of \(\rho\) on Expected Cost}
		\label{fig:corr dis}
	\end{figure}

 \begin{remark}  \label{rem:improvement}
The improvement in the expected cost in Proposition~\ref{iidprop:corr dis} is a nonlinear function of the correlation coefficient \( \rho \). To illustrate the dependency of the improvement in expected cost on \( \rho \) more clearly, we represent it in Figure~\ref{fig:corr dis} by the solid blue lines, which highlight the sensitivity of the improvement in expected cost to changes in the correlation coefficient. The red dashed line represents the variance of the PAA policy. In  Figure~\ref{fig:corr dis}, we choose \( c_o = 1 \), \( c_u = 3\), \( \mu - \mu_d = 0.5 \), \( \sigma^2 = 100 \), and set \( \sigma_d^2 = 1, 10, 100 \) in subfigures (a), (b), and (c), respectively. It is clear that as \( \rho \) increases, the variance of the PAA policy increases and the improvement in expected cost gradually decreases, with no improvement when \( \rho = 1 \). Additionally, as the randomness of the demand, \( \sigma_d^2 \), increases, the improvement in expected cost decreases.
 \end{remark}
 
\begin{proposition}
\label{iidprop:mean dis}
For a fixed sample size \( t \), consider two candidate policies, \( Q^{(1)}(\bm{D}_t) \) and \( Q^{(2)}(\bm{D}_t) \), that follow a two-dimensional normal distribution 
\(
N\left( \begin{pmatrix} \mu_1 \\ \mu_2 \end{pmatrix}, \begin{pmatrix} \sigma^2 & \sigma^2 \\ \sigma^2 & \sigma^2 \end{pmatrix} \right),
\)
where \( \mu_1 \neq \mu_2 \). The demand follows a normal distribution \( N(\mu_d, \sigma_d^2) \), and the demand and candidate policies are assumed to be independent.

Under these conditions, the PAA policy achieves the expected value 
\[
\E[Q^{PAA}(\bm{D}_t)] = \mu_d + \sqrt{\sigma^2 + \sigma_d^2} \Phi^{-1}\Big(\frac{c_u}{c_u + c_o}\Big),
\]
and the smallest expected cost 
\(
J\left(\E[Q^{PAA}(\bm{D}_t)], \sigma^2\right).
\)
The improvement in the expected cost of the PAA policy over the two candidate policies is given by 
\(
J\left(\E[Q^{PAA}(\bm{D}_t)], \sigma^2\right) - J(\mu_1, \sigma^2) \geq 0
\)
and 
\(
J\left(\E[Q^{PAA}(\bm{D}_t)], \sigma^2\right) - J(\mu_2,\sigma^2) \geq 0,
\)
respectively.
\end{proposition}

\begin{remark}
It is counterintuitive that when the order quantity policy has a non-zero variance, the policy that attains the smallest expected cost is not the unbiased estimator of the theoretical optimal constant order quantity, 
\(
Q^{T-opt}=\mu_d + \sigma_d \Phi^{-1}\big(\frac{c_u}{c_u + c_o}\big),
\)
except in the case where \( c_o = c_u \). This observation has practical implications because, in practice, all candidate policies are computed based on the finite data set \( \bm{D}_t \) and are random variables with non-zero variance. 

We note, however, that this observation does not contradict Proposition~\ref{iidprop:PAA converges to the optimal}. As the sample size tends to infinity, the variance of the policy $\sigma^2$ decreases to zero, and the PAA policy $Q^{PAA}(\bm D_t)$ converges to the theoretical optimal constant order quantity.
\end{remark}

Proposition~\ref{iidprop:mean dis} shows that in the case where the candidate policies are perfectly linearly dependent, although the PAA policy cannot achieve a smaller variance through risk diversification, it still attains a smaller expected cost by adjusting the expected value. Based on Propositions~\ref{iidprop:corr dis} and \ref{iidprop:mean dis}, we observe that the improvement of the PAA policy mainly depends on the differences between the candidate policies, whether these differences arise from the expected value or from imperfect linear dependence. In fact, in the following proposition, we demonstrate that the \( \ell_1 \) norm distance between the candidate policies is a determinant of the success of the PAA.

\begin{proposition}
 \label{iidprop:impro from dis}
For a fixed sample size \( t \), consider two candidate policies, \( Q^{(1)}(\bm{D}_t) \) and \( Q^{(2)}(\bm{D}_t) \), which have an \( \ell_1 \) norm distance \( \delta \), i.e., 
\[
\frac{1}{t} \sum_{j=1}^t \left| Q^{(1)}(\bm{D}_{\cancel{j}}) - Q^{(2)}(\bm{D}_{\cancel{j}}) \right| = \delta.
\]
The improvement in expected cost of the PAA policy over the candidate policies is bounded by \( \max(c_o, c_u)\cdot \max(|L|, U) \cdot \delta \).
 \end{proposition}
 
The distance used in Proposition~\ref{iidprop:impro from dis} is the \( \ell_1 \) norm of the difference between two candidate policies. This distance has useful properties, as discussed in detail by \cite{rolle2021various}. Unlike divergences or statistical distances defined for random variables, such as the Wasserstein distance \citep{givens1984class}, which focus solely on the distribution and ignore correlations, the \( \ell_1 \) norm distance accounts for both statistical distances and the correlation between random variables. In the following two remarks, we provide guidance on choosing candidate policies based on Proposition~\ref{iidprop:impro from dis}.

 \begin{remark}  \label{rem:distance}
 From Proposition~~\ref{iidprop:impro from dis}, when the distance between the candidate policies is small, the improvement of the PAA policy is also small. In other words, combining two very similar candidate policies yields little advantage because substantial corrections cannot be made from another similar policy.
 \end{remark}
 
\begin{remark} \label{rem:misspecification}
A large value of the \( \ell_1 \) norm distance often indicates misspecification in the system of candidate policies. If all models used in the candidate policies are well-specified, the distance should not be excessively large. Therefore, candidate policies based on many different models, even if misspecified, are still valuable. As demonstrated in Section~\ref{subsec:simiid} below, a candidate policy with a misspecified model can still improve the performance of a candidate policy with a well-specified model through the PAA.
\end{remark}
 
\subsection{The feature-based demand case} \label{subsec:PAAcorrdem}
In this section, we consider a feature-based demand model, where the observed demand follows an unknown cumulative distribution function (CDF) denoted by \( G(d; \bm{x}) \), with \( \bm{x} \) representing the covariates that influence the demand distribution. There are \( m \) candidate policies, \( Q^{(1)}(\bm{x}; \bm{D}_t), \ldots, Q^{(m)}(\bm{x}; \bm{D}_t) \), each of which is a random function of \( \bm{x} \) and may be generated using either model-based or data-driven approaches. Similar to the i.i.d. demand case, we omit the parameters \( \hat{\bm{\theta}}(\bm{D}_t) \), \( \tilde{\bm{\theta}}(\bm{D}_t) \), and \( \tilde{\bm{q}}(\bm{D}_t) \) in the candidate policies to simplify the notation. For a given \( \bm{x} \), we assume the existence of an unknown joint CDF \( F(Q^{(1)}(\bm{x}; \bm{D}_t), \ldots, Q^{(m)}(\bm{x}; \bm{D}_t)) \) for the candidate policies.

Although this assumption is similar to the i.i.d. case when the covariate \( \bm{x} \) is fixed, challenges arise when there are insufficient samples for specific values of \( \bm{x} \). In such cases, we cannot directly construct an empirical version of the objective in $(PAA)$ with conditional expected cost. To address this issue, we propose an alternative approach for implementing the PAA in the feature-based demand setting by considering the total cost across all available samples. Specifically, we assume the existence of the joint CDFs \( G(d, \bm{x}) \) and \( F(Q^{(1)}(\bm{x}; \bm{D}_t), \ldots, Q^{(m)}(\bm{x}; \bm{D}_t), \bm{x}) \).

\begin{procedure}\label{proc:fb}{\bf [Policy averaging approach for feature-based demand case]}\\
1. Define the reduced data set without $(d_j,\bm x_j)$ as 
\[\bm D_{\cancel{j}}=\{(d_1,\bm x_1),\ldots,(d_{j-1},\bm x_{j-1}), (d_{j+1},\bm x_{j+1}),\ldots,(d_t,\bm x_t)\}.\]
2. Compute the $m$ candidate policies over  $\bm D_{\cancel{j}}$, $Q^{(1)}(\bm x;\bm D_{\cancel{j}})$, $\ldots$, $Q^{(m)}(\bm x;\bm D_{\cancel{j}})$.\\
3. Determine the optimal weights $\omega_1^*,\ldots,\omega_m^*$ by minimizing the average  cost over unused data, \begin{align*}
\min_{\{\omega_1,\ldots,\omega_m\}}~\textstyle\sum_{j=1}^t C(\sum_{i=1}^{m}\omega_i Q^{(i)}(\bm x_j;\bm D_{\cancel{j}})-d_j).
\end{align*}
\end{procedure}

By adding the constraints for weights, we have the following optimization problem $(PAA_{FB})$,
\begin{align*}
(PAA_{FB}) \quad \min_{\{\omega_1, \ldots, \omega_m\}}~ & \frac{c_o}{t} \sum_{j=1}^t \max (Q(\x_j;\bm D_{\cancel{j}}) - d_j, 0) + \frac{c_u}{t} \sum_{j=1}^t \max (d_j - Q(\x_j;\bm D_{\cancel{j}}), 0) \\
\text{s.t.} \quad & Q(\x_j;\bm D_{\cancel{j}}) = \sum_{i=1}^m \omega_i Q^{(i)}(\x_j;\bm D_{\cancel{j}} ), \\
& L \leq \omega_i \leq U, \\
& \sum_{i=1}^m \omega_i = 1,
\end{align*}
where $L\leq 0$ and $U\geq 1$. Similar to $(PAA_{iid})$, the optimization problem $(PAA_{FB})$ is still convex. We denote the obtained PAA policy on data set $\bm D_t$ as $Q^{PAA}(\bm x; \bm D_t) =\sum_{i=1}^{m}\omega_i^* Q^{(i)}(\bm x; \bm D_t)$. 

\begin{remark}
Although the procedure for the feature-based case appear quite similar to the i.i.d. case, there are substantial differences. First, in the i.i.d. case, the PAA assigns different weights to different random variables, whereas in the feature-based case, it assigns weights to different random functions. Second, in the feature-based case, there is additional randomness arising from the covariates, which plays a crucial role in determining the optimal weights.
\end{remark}

To simplify the theoretical discussion for the feature-based case, we assume that the covariate \( x \) is a one-dimensional random variable. Following \cite{ban2019big}, the \( i \)th candidate policy is represented as 
\(
Q^{(i)}(x; \bm{q}^{(i)}(\bm{D}_{\cancel{j}})) = \sum_{k=0}^p q_{k}^{(i)}(\bm{D}_{\cancel{j}}) x^k,
\)
which is a polynomial function of the covariate \( x \). Based on Taylor expansion, the polynomial form is a suitable approximation for a general demand model (\citealp{ban2019big}). Under these assumptions, the candidate policies are uniquely determined by the random vectors \( \bm{q}^{(1)}(\bm{D}_{\cancel{j}}), \ldots, \bm{q}^{(m)}(\bm{D}_{\cancel{j}}) \). The PAA policy is then characterized by the vector
\(
\bm{q}^*(\bm{D}_{\cancel{j}}) = \sum_{i=1}^{m} \omega_i^* \bm{q}^{(i)}(\bm{D}_{\cancel{j}}).
\)
We can now state the following propositions, which provide the asymptotic and finite sample properties of the PAA in the feature-dependent case. 

\begin{proposition}  
\label{fbprop:PAA is the best}
When the demand and covariate data are obtained from a joint CDF and the sample size tends to infinity (\( t \to \infty \)), the PAA policy can achieve an expected cost that is no greater than the expected total cost of any candidate policy.
\end{proposition}
 
\begin{proposition}
\label{fbprop:PAA converges to the optimal}
Under the following assumptions:
\begin{itemize}[itemsep=-1pt]
    \item The demand and covariate data are obtained from a joint CDF;
    \item The random parameter vector of the \(i\)th candidate policy, \(\bm{q}^{(i)}(\bm{D}_t)\), converges to a deterministic vector \(\bar{\bm{q}}^{(i)}\) as the sample size \(t \to \infty\), with a standard deviation vector of \(\frac{\bm{M}_i}{\sqrt{t}}\), where \(\bm{M}_i\) is a constant vector, for \(i = 1, \dots, m\);
    \item \(\{\bar{\bm{q}}^{(1)}, \dots, \bar{\bm{q}}^{(m)}\}\) form a system of vectors containing \(p+1\) independent vectors;
    \item The lower bound \(L\) and upper bound \(U\) for the data set \(\bm{D}_t\) approach negative and positive infinity, respectively, at rates slower than \(-\sqrt{t}\) and \(\sqrt{t}\);
\end{itemize}
then, the PAA policy \( Q^{PAA}(x; \bm{D}_t) \) converges to the optimal order quantity in the function space of \( p \)-degree polynomials as the sample size \( t \to \infty \).
\end{proposition}

\begin{proposition}\label{fbprop:impro from dis}
For a fixed sample size \( t \), consider two candidate policies, \( Q^{(1)}(x;\bm{D}_{\cancel{j}}) \) and \( Q^{(2)}(x;\bm{D}_{\cancel{j}}) \), which have an \( \ell_1 \) norm distance \( \delta \), i.e., 
\[
\frac{1}{t} \sum_{j=1}^t \left| Q^{(1)}(x_j;\bm{D}_{\cancel{j}}) - Q^{(2)}(x_j;\bm{D}_{\cancel{j}}) \right| = \delta.
\]
The improvement in expected cost of the PAA policy over the candidate policies is bounded by \( \max(c_o, c_u)\cdot \max(|L|,U)\cdot \delta \).
\end{proposition}

\begin{remark}
Proposition \ref{fbprop:PAA converges to the optimal} can be extended to a more general version. When the candidate policies converge to general nonlinear functions of the covariate, rather than deterministic polynomial functions, the PAA converges to an optimal policy in the linear space spanned by the general nonlinear functions.
\end{remark}

Next, we focus on the impact of the PAA policy on the overfitting of candidate policies. As computational power increases, many data-driven policies have become popular, but they often suffer from overfitting. Suppose there are two overfitting candidate policies of polynomial type:
\begin{align*}
& Q^{(1)}(x;\bm{D}_{t}) = \sum_{k=0}^{p} q_k^{(1)}(\bm{D}_{t}) x^k + \sum_{s=p+1}^{p^{(1)}} q_s^{(1)}(\bm{D}_{t}) x^s \triangleq \beta^{(1)}(x;\bm{D}_{t}) + \delta^{(1)}(x;\bm{D}_{t}), \\
& Q^{(2)}(x;\bm{D}_{t}) = \sum_{k=0}^{p} q_k^{(2)}(\bm{D}_{t}) x^k + \sum_{s=p+1}^{p^{(2)}} q_s^{(2)}(\bm{D}_{t}) x^s \triangleq \beta^{(2)}(x;\bm{D}_{t}) + \delta^{(2)}(x;\bm{D}_{t}),
\end{align*}
where \( p \) is the correct order of the polynomial function, while \( p^{(1)} > p \) and \( p^{(2)} > p \) are the orders of the polynomial functions in the candidate policies. This implies that both candidate policies are overfitting. The terms \( \beta^{(1)}(x;\bm{D}_{t}) \) and \( \beta^{(2)}(x;\bm{D}_{t}) \) represent the correct parts, while \( \delta^{(1)}(x;\bm{D}_{t}) \) and \( \delta^{(2)}(x;\bm{D}_{t}) \) represent the overfitting parts. For the overfitting policy \( Q^{(i)}(x;\bm{D}_{\cancel{j}}) \), $i=1,2$, we often have
\[
    \mathbb{E}[Q^{(i)}(x;\bm{D}_{t})] = \mathbb{E}[Q^{T-\text{opt}}(x)], \quad \text{Var}(Q^{(i)}(x;\bm{D}_{t})) > \text{Var}(\beta^{(i)}(x;\bm{D}_{t})),
\]
where \( \text{Var}(Q^{(i)}(x;\bm{D}_{t})) - \text{Var}(\beta^{(i)}(x;\bm{D}_{t})) \) measures the degree of overfitting.

\begin{proposition}\label{fbprop:overfitting}
For a fixed sample size \( t \), consider two candidate policies, \( Q^{(1)}(x;\bm{D}_t) \) and \( Q^{(2)}(x;\bm{D}_t) \), that follow a two-dimensional normal distribution 
\(N\left( \begin{pmatrix} \mu \\ \mu \end{pmatrix}, \begin{pmatrix} \sigma^2 & \rho \sigma^2 \\ \rho \sigma^2 & \sigma^2 \end{pmatrix} \right)
\), where the correlation coefficient \( \rho < 1 \). \(\text{Var}(\beta^{(1)}(x;\bm{D}_t))=\text{Var}(\beta^{(2)}(x;\bm{D}_t))=\sigma_{\beta}^2\). The demand follows a normal distribution \( N(\mu_d, \sigma_d^2) \), and the demand and candidate policies are assumed to be independent. Under these conditions, the optimal weight on the first candidate policy is given by
\(
\omega_1^* = \frac{1}{2}.
\)

Furthermore, when the overfitting candidate policies satisfy the following conditions:
\begin{align*}
\sigma^2 > \sigma_{\beta}^2,\quad \quad \rho \sigma^2 < \rho_{\beta} \sigma_{\beta}^2, 
\end{align*}
where \( \rho_{\beta} \) is the correlation coefficient between \( \beta^{(1)}(x;\bm{D}_t) \) and \( \beta^{(2)}(x;\bm{D}_t) \), the following inequality holds:
\begin{align*}
    & \sigma^2 - \sigma_{\beta}^2 > \text{Var}(Q^{\text{PAA}}(x;\bm{D}_t)) - \text{Var}(\omega_1^* \beta^{(1)}(x;\bm{D}_t) + (1-\omega_1^*) \beta^{(2)}(x;\bm{D}_t)).
\end{align*}
\end{proposition}

Proposition \ref{fbprop:overfitting} shows that when combining two overfitting candidate policies, the degree of overfitting in the PAA policy can be reduced. The first condition for this result is that the candidate policies are overfitting. The second condition requires that the correlation coefficient between the candidate policies is small. In fact, when the second candidate policy is not overfitting and its mean is close to the mean of the first overfitting candidate policy, i.e., \( \delta^{(2)}(x;\bm{D}_t) = 0 \) and \( \mathbb{E}[Q^{(2)}(x;\bm{D}_t)] \approx \mathbb{E}[Q^{(1)}(x;\bm{D}_t)] \), the second inequality becomes much easier to satisfy. Therefore, it is beneficial to include well-performing model-based candidate policies which are typically not overfitting, in order to reduce the overfitting of more complex data-driven policies.

 \section{Simulation Study}  \label{sec:simulation}
 
In order to compare the performance of the PAA with previous solution approaches to the newsvendor problem, we use simulation with randomly generated data sets. Section~\ref{subsec:simiid} conducts a simulation study for the case of i.i.d. demand. We consider three configurations of candidate policies, and identify a general relationship between the distance between them and the performance of the PAA.  Section~\ref{subsec:simfeature} conducts a simulation study for the feature-based demand case. Here, we mainly focus on the overfitting issue.

\subsection{The i.i.d. demand case}\label{subsec:simiid}
   
We first simulate $t+10,~t=20,24,\ldots,200$ data points from the normal distribution with mean $60$ and standard deviation $10$, and set $c_o=1$, $c_u=3$. We use the first $t$ data points to compute the candidate policies as well as the optimal weights of the PAA policy, following Procedure \ref{proc:iid}. The last 10 data points, $t+1,\ldots,t+10$, are omitted to serve as out of sample data for comparing the performance of the different policies. This procedure repeats 1000 times. The performance criteria used are average order quantity and average out of sample cost.

We consider candidate policies generated by five different distributional assumptions that are summarized in Table~\ref{tab:summary of iid policy}. In the fourth column, the description``Mis-/Well-specified'' follows Definitions~\ref{def:well_specified} and \ref{def:mis_specified}. In the last column, ``Over-/Under-estimated'' implies that the expected order quantity of the candidate policy is larger/smaller than the theoretical optimal order quantity $Q^{T-opt}=60+10\Phi^{-1}(\frac{c_u}{c_o+c_u})=66.7449$. 

\begin{table}[!h]
\centering
\footnotesize
\begin{tabular}{ccccc}
\toprule
\textbf{Policy} & \textbf{Distribution} & \textbf{ETO/IEO} & \textbf{Mis-/Well-specified} & \textbf{ Over-/Under-estimated}
\\
\hline
$Q^{(1)}$ & Inverse-Gamma      & ETO & Mis-specified & Under-estimated \\ 
$Q^{(2)}$ & Student $t$   & ETO   & Mis-specified & Under-estimated\\
$Q^{(3)}$ &    Exponential       & ETO  & Mis-specified & Over-estimated\\ 
$Q^{(4)}$ &  Chi-square         & ETO   & Mis-specified & Over-estimated\\
$Q^{(5)}$ & Normal   & IEO   & Well-specified & Unbiased\\
\bottomrule\end{tabular}
\caption{Five Candidate Policies Generated by Different Distributional Assumptions}
\label{tab:summary of iid policy}
\end{table}

\begin{figure}[htbp]
		\centering
		\begin{subfigure}[b]{0.45\textwidth}
			\centering
			\includegraphics[height=6.3cm]{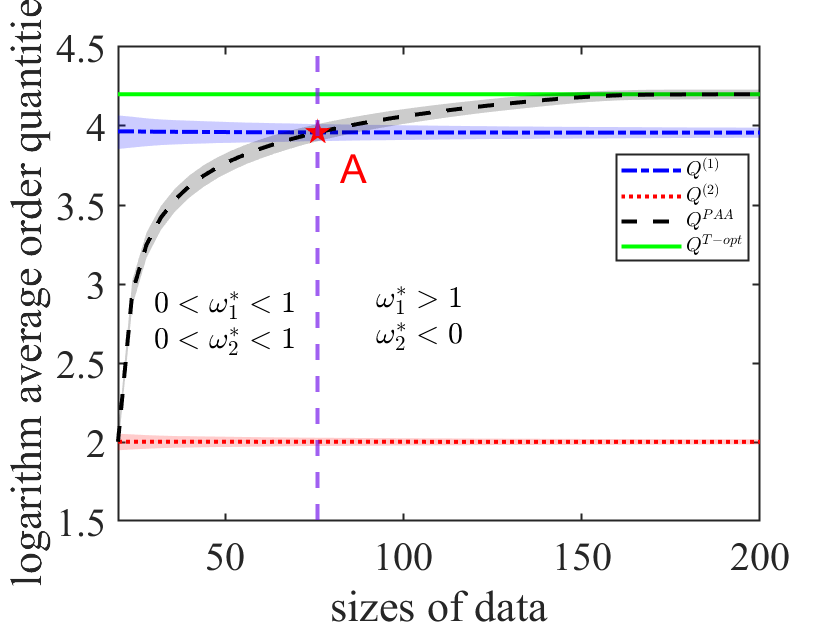} 
			\caption{Average Order Quantities}
			\label{iidfig:twomis:subfig:a}
		\end{subfigure}
		\hfill
		\begin{subfigure}[b]{0.45\textwidth}
			\centering
		\includegraphics[height=6.3cm]{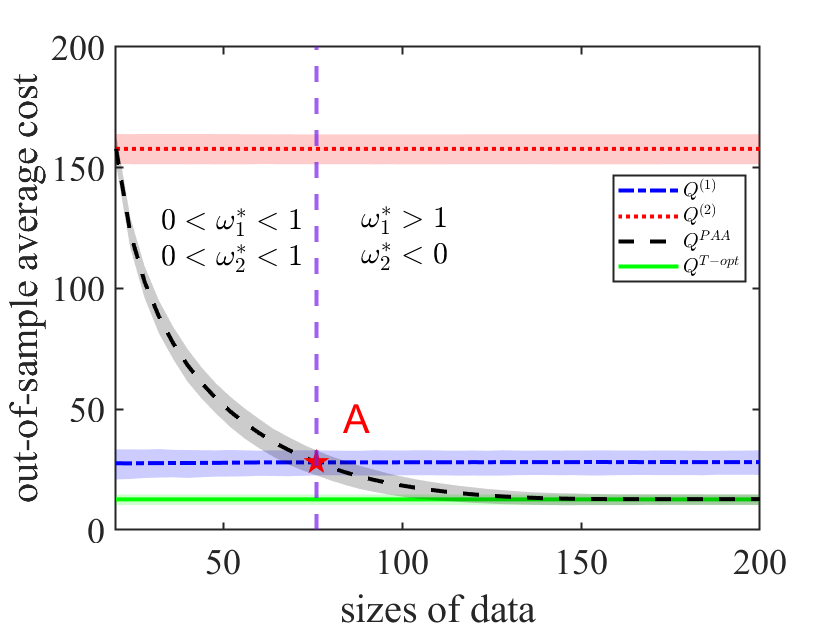} 
			\caption{Out-of-sample Average Cost}
			\label{iidfig:twomis:subfig:b}	
		\end{subfigure}
		\caption{PAA for Two Mis-specified and Underestimated Candidate Policies}
		\label{iidfig:twomis}
	\end{figure}

We first choose candidate policies \(Q^{(1)}\) and \(Q^{(2)}\)  to generate the PAA policy, then we compute $Q^{PAA}=\omega^*_1Q^{(1)}+\omega^*_2Q^{(2)} $ by using $L = -\frac{\log(n)}{15}$ and $U=1+\frac{\log(n)}{15}$. We observe that both these candidate policies emanate from wrong distributional assumptions, and both are underestimated. Hence, we study the performance of the PAA policy where the candidate policies consistently fall short of the optimal order quantity. Figure~\ref{iidfig:twomis} shows the average order quantities and out-of-sample performance for the two candidate policies and the PAA policy.

Figure~\ref{iidfig:twomis}(a) reports the (logarithm of) average order quantities and the shaded areas around the lines represent their $95\% $ confidence interval, respectively. Figure~\ref{iidfig:twomis}(b) reports the out-of-sample average cost and the shaded areas denotes their inter-quartile range. The results demonstrate clearly better out-of-sample performance by the PAA policy than the candidate policies. Both candidate policies are derived from incorrect distributional assumptions, hence neither policy converges to the theoretical optimal policy $Q^{T-opt}$. Statistically speaking, the estimated inverse gamma distribution is closer to the true normal distribution than the estimated student $t$ distribution as measured by Kullback–Leibler divergence. Therefore, we expect that $Q^{(1)}$ provides estimates closer to the theoretical optimal order quantity than those from $Q^{(2)}$. This expectation is indeed reflected in Figure~\ref{iidfig:twomis}, which shows that $Q^{(1)}$ is closer to the theoretical optimal policy compared to $Q^{(2)}$, and that $Q^{(1)}$ exhibits better out-of-sample performance than $Q^{(2)}$.

Thus, by collecting wisdom from two candidate policies, the PAA policy converges to the theoretical optimal order quantity, consistent with the theoretical result in Proposition \ref{iidprop:PAA is the best}. Furthermore, as shown in Figure~\ref{iidfig:twomis}(b), the PAA policy also provides the best out-of-sample performance. Additional insight is provided by Point~A in Figure ~\ref{iidfig:twomis}, where $\omega_1^* = 1$ and $\omega_2^* = 0$. This point shows where the PAA policy not only surpasses the performance of the superior candidate policy $Q^{(1)}$ in converging to the optimal order quantity, but also achieves improved out-of-sample performance. To the left of Point A,  $0<\omega_1^* < 1$ and $0< \omega_2^* < 1$. Conversely, to the right of Point A, $\omega_1^* >1$ and $\omega_2^* < 0$. Essentially, as the size of the data increases, the PAA policy progressively allocates more weight to the more accurate policy $Q^{(1)}$, while also leveraging the information in $Q^{(2)}$ through negative weights on $\omega_2^*$. This result is intuitive: by harnessing the collective insights from the candidate policies and meanwhile diversifying the model risk, the PAA policy outperforms each candidate policy and, more impressively, converges to the optimal solution.

\begin{figure}[htbp]
    \centering
     \hspace*{-0.1\textwidth} 
    \begin{minipage}{\textwidth}
        \centering
        \begin{subfigure}[b]{0.45\textwidth}
            \centering
            \includegraphics[height=6.6cm]{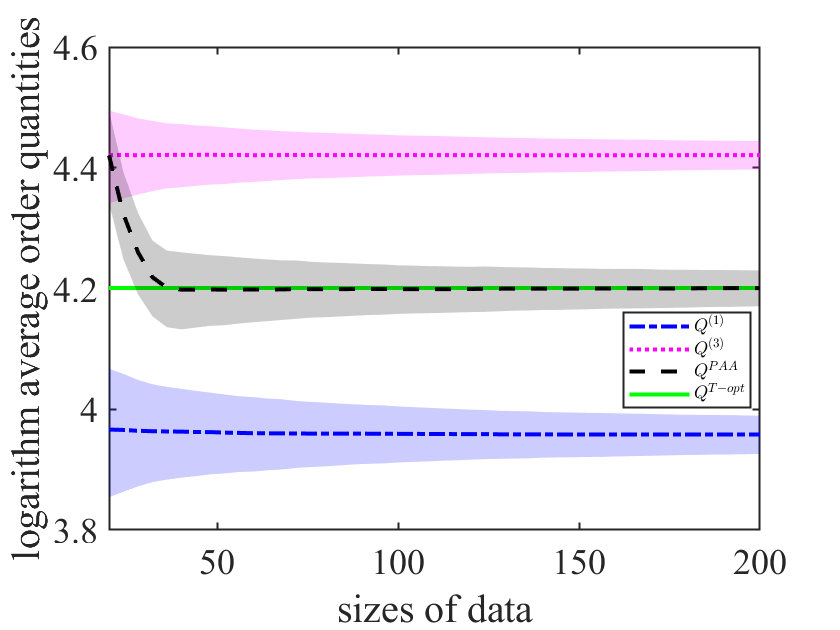}
            \caption{Average Order Quantities}
            
        \end{subfigure}
        \hfill
        \begin{subfigure}[b]{0.45\textwidth}
            \centering
            \includegraphics[height=6.6cm]{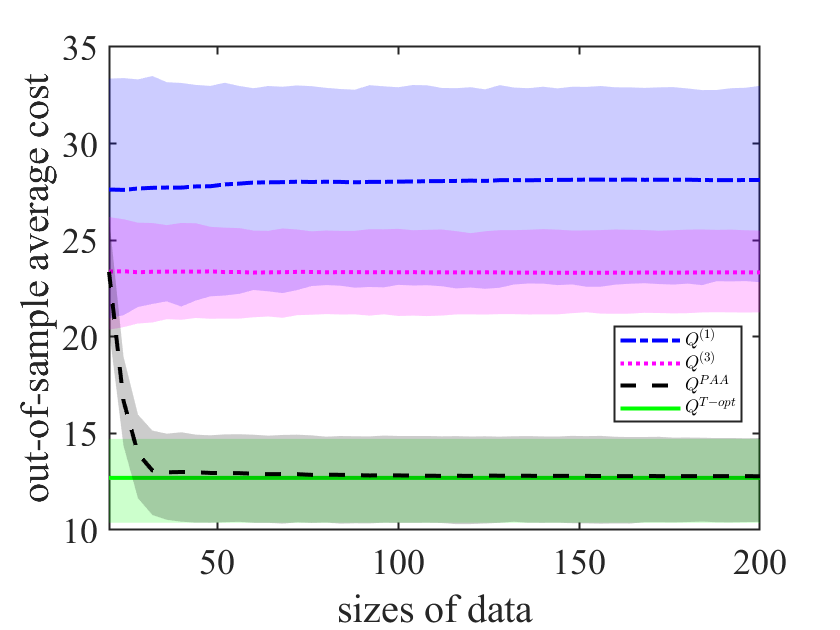}
            \caption{Out-of-sample Average Cost}
            
        \end{subfigure}
    \end{minipage}
    \caption{PAA for Conservative and Aggressive Candidate Policies}
    \label{iidfig:aggressive and conservative}
\end{figure}

Our second study considers two different mis-specified candidate policies, the original conservative $Q^{(1)}$ and an aggressive $Q^{(3)}$, to investigate the performance of the PAA  in aggregating two different types of estimators. Figure~\ref{iidfig:aggressive and conservative}(a) reports the average order quantity, where the shaded areas denotes their 95\% confidence interval. Figure~\ref{iidfig:aggressive and conservative}(b) reports the out-of-sample average cost and the shaded areas denotes their inter-quartile range. Similar to Figure~\ref{iidfig:twomis}, the PAA policy  converges to the optimal order quantity, as shown by Figure~\ref{iidfig:aggressive and conservative}(a), and also exhibits the best out-of-sample performance, as shown by Figure~\ref{iidfig:aggressive and conservative}(b). Moreover, in comparison to Figure~\ref{iidfig:twomis}, the convergence of the PAA policy is faster when integrating divergent policy estimates. 

\begin{figure}[htbp]
    \centering
     \hspace*{-0.1\textwidth} 
    \begin{minipage}{\textwidth}
        \centering
        \begin{subfigure}[b]{0.48\textwidth}
            \centering
            \includegraphics[height=6.3cm]{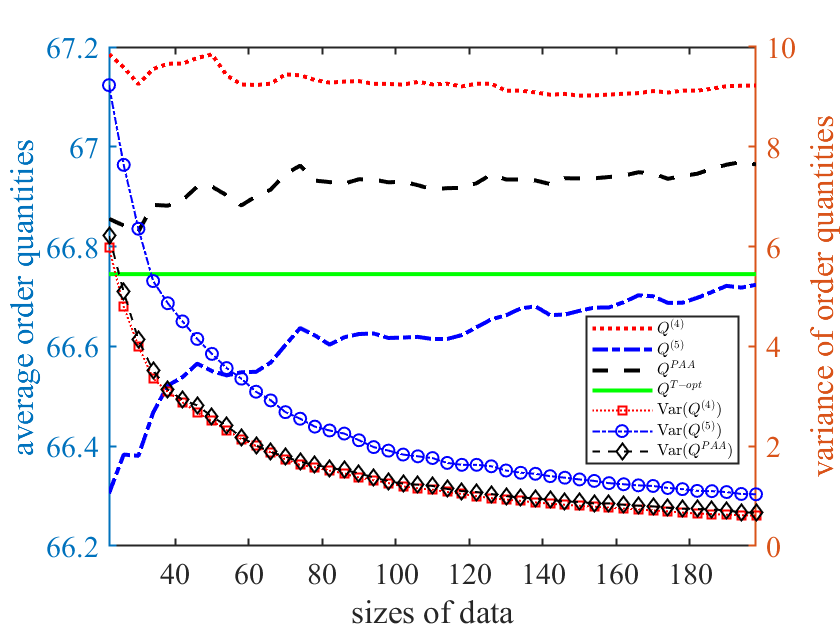}
            \caption{Average Order Quantities}
           
        \end{subfigure}
        \hfill
        \begin{subfigure}[b]{0.42\textwidth}
            \centering
            \includegraphics[height=6.3cm]{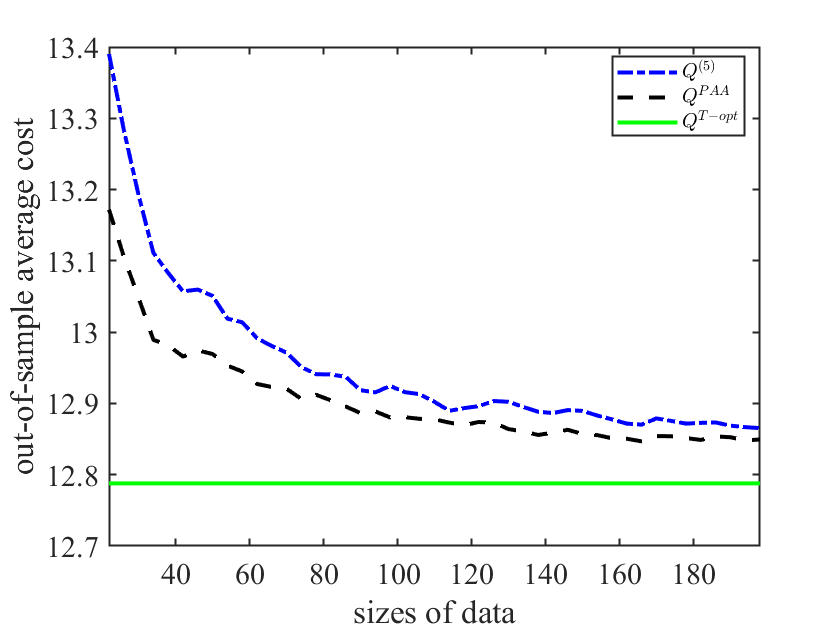}
            \caption{Out-of-sample Average Cost}
            
        \end{subfigure}
    \end{minipage}
    \caption{PAA for Well-specified and Mis-specified Candidate Policies}
    \label{iidfig:onewellonemisorder}
\end{figure}

For our third study, we apply a well-specified policy $Q^{(5)}$ alongside a mis-specified one $Q^{(4)}$ as our two candidate policies. This extension allows us to explore whether our PAA approach can further enhance the performance of the well-specified policy. This example sheds light on the effectiveness of the PAA in optimizing decision-making under uncertainty, even when one of the candidate policies is already well-specified to the underlying stochastic process. The results appear in 
Figure~\ref{iidfig:onewellonemisorder}.

Figure~\ref{iidfig:onewellonemisorder}(a) presents the average order quantities and Figure~\ref{iidfig:onewellonemisorder}(b) reports the out-of-sample average cost, for both the candidate policies and the PAA policies.  Figure~\ref{iidfig:onewellonemisorder}(b) clearly demonstrates that the PAA improves the performance of the well-specified policy $Q^{(5)}$ by significantly reducing the out-of-sample average cost, as evidenced by the transition from the blue dot-dash line to the black dashed line.  Figure~\ref{iidfig:onewellonemisorder}(a) complements this observation by plotting the variance of each policy, as viewed from the right-hand-side y-axis, which shows the source of improvement. The reduction in out-of-sample average cost is primarily attributable to the variance reduction achieved by incorporating the mis-specified policy $Q^{(4)}$. Importantly, both the variance reduction and the improved out-of-sample average cost come at a cost, which is a trade-off in the accuracy of the average mean of the order quantity. The PAA policy deviates  slightly more from the optimal order quantity than the original well-specified policy $Q^{(5)}$. Overall, the PAA improves the performance of the well-specified policy in  reducing the variance and out-of-sample average cost.

\begin{remark}\label{rem:pattern}
In the above analysis, we provide a detailed study of the performance of the PAA by combining two distinct types of candidate policies: a pair of underestimated policies,\(\{Q^{(1)},Q^{(2)}\}\);  a combination of an underestimated and an overestimated policy, \(\{Q^{(1)},Q^{(3)}\}\); and a pair consisting of a well-specified and a mis-specified policy \(\{Q^{(4)},Q^{(5)}\}\). We observe that these simulation  results show a  consistent pattern: the greater the divergence between the two candidate policies, the greater the performance improvement by the PAA. This is consistent with the results in Proposition~ \ref{iidprop:impro from dis}.
\end{remark} 

To investigate this observed trend in Remark \ref{rem:pattern}, we fix the data size at $n=120$ and explore the overall performance of the PAA by examining all the 10 possible combinations of the five candidate policies listed in Table~\ref{tab:summary of iid policy}. Based on Proposition~\ref{iidprop:impro from dis}, we use the \(\ell_1\) norm as a metric to quantify the divergence between two candidate policies. The objective of our investigation is to understand whether there is a positive correlation between the magnitude of this distance and the extent of improvement achieved by the PAA.

\begin{figure}[htbp]
        \centering
        \includegraphics[width=0.6\textwidth]{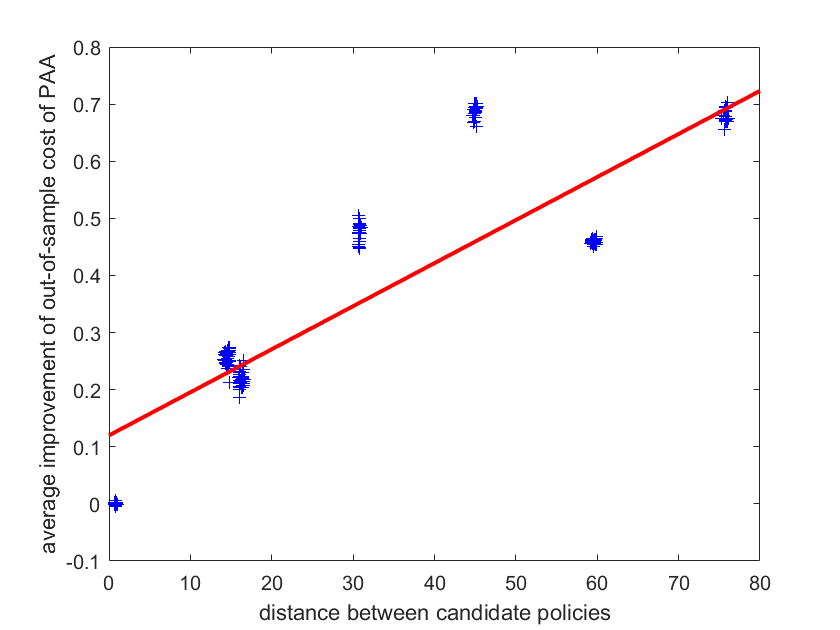}
           
            \caption{Distance between Candidate Policies with Relative Improvement in Cost}
            \label{iidfig:distance and improve}   
    \end{figure}

To represent the variation in both distance and improvement more completely, we divide the 1000 sample paths of $t=120$ into 20 groups of 50 sample paths. For the 50 sample paths within each group, we first compute the average distance between the candidate policies and the average cost for both the PAA and all individual candidate policies. Then, the relative improvement in cost is calculated as the average ratio of the PAA cost reduction to the costs of the two candidate policies. Then, in the (distance, relative improvement) space shown in Figure~\ref{iidfig:distance and improve}, we plot a total of 200 points. The results in Figure~\ref{iidfig:distance and improve} demonstrate a clear positive relationship
between improvement of cost and distance of candidate policies, consistent with the theoretical finding in Proposition \ref{iidprop:impro from dis}. More specifically, the $R^2$ value of the regression between the relative improvement and the distance is $0.74$ and the $p$-value of the slope coefficient is less than $1\times10^{-60}$.

\subsection{The feature-dependent demand case}  \label{subsec:simfeature}

We assume that the demand follows a normal distribution \( N(\mu(x), (\sigma(x))^2) \), where the mean and standard deviation are characterized by the following functions:
\[
\mu(x) = 80 e^{-10 (x + 0.5)^2} + 30, \quad \sigma(x) = 5 e^{-8 (x - 0.5)^2}.
\]
Furthermore, we assume that the covariate \( x \) follows a uniform distribution over the interval \( [-1, 1] \), with \( c_o = 1 \) and \( c_u = 3\). Subsequently, we generate \( 2t \) data points for each \( t = 20, 30, \dots, 100, 200, 300, 400, 500 \) based on this setup. We use the first \( t \) data points to compute the candidate policies and the optimal weights of the PAA policy, following Procedure \ref{proc:fb}. The remaining \( t \) data points, \( t+1, \dots, 2t \), are omitted and used as out-of-sample data to evaluate the performance of the different approaches. This procedure repeats 1000 times. The performance criteria used are the average order quantity and the average out-of-sample cost. Unlike in the i.i.d. case, the main focus here is on the role of the PAA approach in controlling overfitting issues.

We consider three candidate policies generated using different methodological assumptions, as summarized in Table~\ref{tab:summary_of_candidate_policies}. The PAA policy is then computed as
\(
Q^{\text{PAA}}(x) = \omega^*_1 Q^{(1)}(x) + \omega^*_2 Q^{(2)}(x) + \omega^*_3 Q^{(3)}(x),
\)
where \( L = -\log(n) \) and \( U = 1 + \log(n) \). Concurrently, we provide the theoretical optimal order quantity,
\(
Q^{T-\text{opt}}(x) = \mu(x) + \sigma(x) \Phi^{-1}\big( \frac{c_u}{c_o + c_u} \big).
\)

\begin{table}[!h]
\centering
\footnotesize
\renewcommand{\arraystretch}{1.2} 
\begin{tabular}{p{1.3cm}p{3.2cm}p{2cm}p{8cm}}
\toprule
\textbf{Policy} & \textbf{Model-based /Data-driven} & \textbf{Overfitting} & \textbf{Key Features} \\
\hline
$Q^{(1)}(x)$ & Model-based & No & \text{Demand follows} $N(\mu^{(1)}(x), (\sigma^{(1)}(x))^2)$, where both $\mu^{(1)}(x)$ and $\sigma^{(1)}(x)$ are linear functions of $x$. \\
$Q^{(2)}(x)$ & Data-driven & Yes & Order quantity is a nonlinear function of $x$, which is constructed by Support Vector Machine (SVM) with Radial basis function (RBF) kernel and penalty parameter is 100. \\
$Q^{(3)}(x)$ & Data-driven & Yes & Order quantity is a nonlinear function of $x$, which is constructed by Neural Networks with two hidden layers (20 and 10 neurons) and ReLU activation. \\
\bottomrule
\end{tabular}
\caption{Summary of Candidate Policies}
\label{tab:summary_of_candidate_policies}
\end{table}

\begin{figure}[h]
    \centering
    \begin{minipage}[t]{0.45\textwidth}
        \centering
        \includegraphics[width=\linewidth]{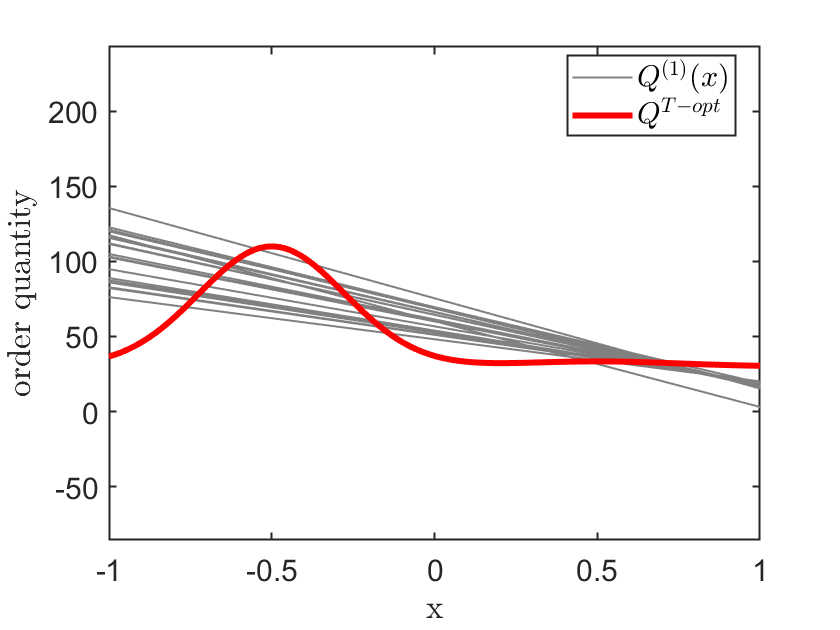} 
        \subcaption{ETO Linear} 
        \label{fig:fborderquantity:a}
    \end{minipage}
    \hfill
    \begin{minipage}[t]{0.45\textwidth}
        \centering
        \includegraphics[width=\linewidth]{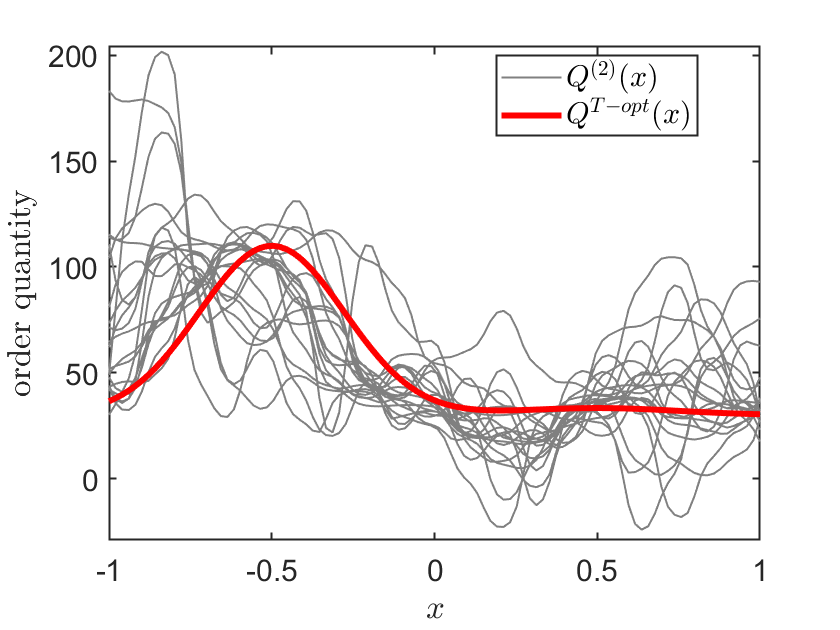} 
        \subcaption{SVM} 
        \label{fig:fborderquantity:b}
    \end{minipage}
    
    \vspace{0.5cm} 
    \begin{minipage}[t]{0.45\textwidth}
        \centering
        \includegraphics[width=\linewidth]{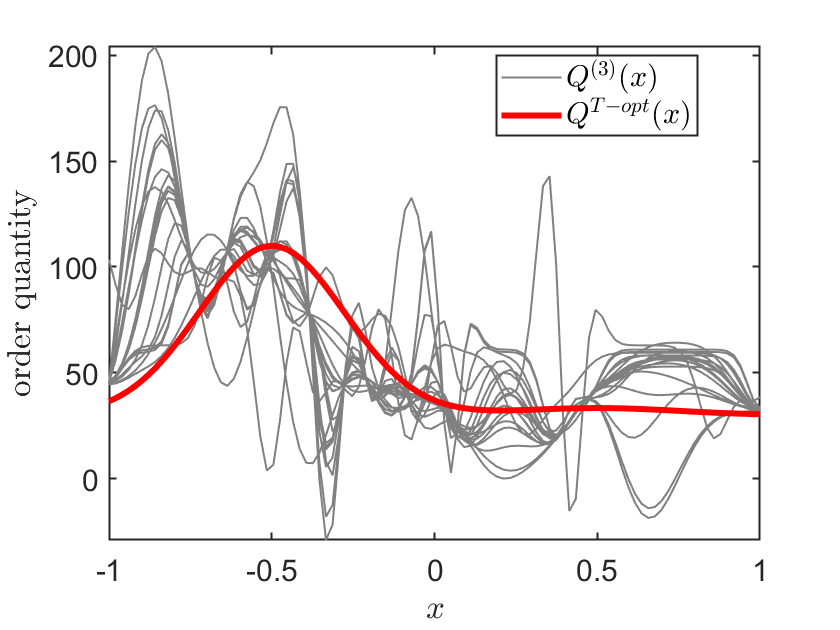} 
        \subcaption{Neural Network} 
        \label{fig:fborderquantity:c}
    \end{minipage}
    \hfill
    \begin{minipage}[t]{0.45\textwidth}
        \centering
        \includegraphics[width=\linewidth]{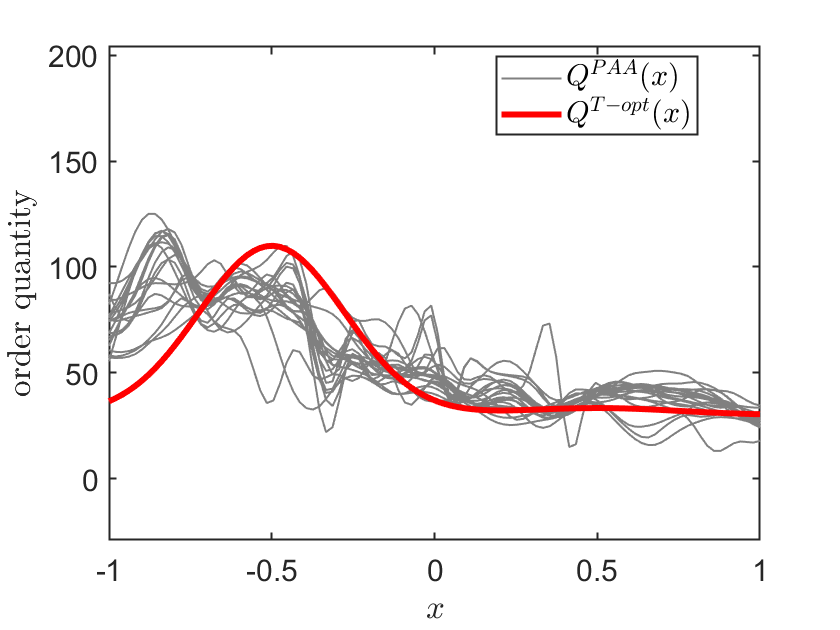} 
        \subcaption{PAA} 
        \label{fig:fborderquantity:d}
    \end{minipage}

    \caption{Analysis of Order Quantity Across Different Approaches} 
    \label{fig:fborderquantity}
\end{figure}

Based on the dataset with \( t = 20 \), we can select any \( t - 1 = 19 \) data points to construct the candidate policies. Therefore, for each candidate policy, we obtain 20 estimated functions. For the first simulated dataset with \( t = 20 \) (out of 1000 repetitions), we compute the 20 estimated functions for the three candidate policies as well as the PAA policy. The 20 estimated functions for each policy are represented by the thin gray curves, while the theoretical order quantity is shown by the thick red curve, as depicted in Figure~\ref{fig:fborderquantity}.

In Figure~\ref{fig:fborderquantity:a}, the ETO Linear policy exhibits a functional form that deviates from the theoretically optimal order quantity \( Q^{T-opt}(x) \). This discrepancy arises from a misspecified model structure that fails to capture the correct relationship between the feature \( x \) and the order quantity. Despite this misspecification, the ETO Linear policy produces highly stable results with minimal variability. While this stability is achieved at the cost of being further from the optimal solution, it helps the PAA policy mitigate the overfitting issues of the data-driven policies. Figures~\ref{fig:fborderquantity:b} and \ref{fig:fborderquantity:c} demonstrate the performance of the SVM and Neural Network policies, respectively. These data-driven policies exhibit nonlinear functional forms. However, the gray curves in both subplots reveal significant variability, which can be interpreted as a symptom of overfitting. The large fluctuations in the order quantities indicate that the data-driven models are overly sensitive to the training data, resulting in less stable performance. In Figure~\ref{fig:fborderquantity:d}, the PAA  integrates the strengths of the candidate policies, yielding order quantities that are both more accurate and more stable. Specifically, the PAA improves upon \( Q^{(1)}(x) \) by incorporating information from the more complex data-driven policies \( Q^{(2)}(x) \) and \( Q^{(3)}(x) \). Conversely, it addresses the overfitting issues in \( Q^{(2)}(x) \) and \( Q^{(3)}(x) \) by leveraging the stability of \( Q^{(1)}(x) \). This complementary interaction between model-based and data-driven approaches enables the PAA to achieve a balance, resulting in order quantities that are closer to the theoretical optimum \( Q^{T-opt}(x) \), while also exhibiting reduced variability.

Table~\ref{tab:fdcomparisons} summarizes the performance of different policies in terms of average costs across the training, testing, and out-of-sample sets in the experiment. The column labeled \( t \) represents the size of the training set. As shown, the two data-driven policies, particularly the neural network policy, perform exceptionally well on the training set, significantly outperforming the theoretical optimal policy. However, on both the testing and out-of-sample sets, the neural network policy's average cost more than doubles compared to its performance on the training set when the sample size is small, highlighting a pronounced overfitting issue. Similarly, the SVM policy also shows a substantial increase in cost — exceeding 100\% — further emphasizing the challenge of overfitting in data-driven policies with limited training data.
In contrast, the PAA policy, which utilizes reweighting, demonstrates superior performance compared to all candidate policies on both the testing and out-of-sample sets. By effectively mitigating the overfitting problem, the PAA policy achieves better performance. Furthermore, as the sample size increases, the out-of-sample average cost of the PAA policy gradually converges towards the theoretical optimal policy.

\begin{table}[h]
    \centering
    \footnotesize 
    \setlength{\tabcolsep}{1.5pt} 
    \renewcommand{\arraystretch}{1.2} 
\begin{tabular}{cccccccccccccccc}
\toprule
\multirow{2}[4]{*}{$t$} & \multicolumn{3}{c}{Training Perf.} &       & \multicolumn{4}{c}{Testing Perf.} &       & \multicolumn{4}{c}{Out-of-sample Perf.} &       & Perf. \\
\cmidrule{2-4}\cmidrule{6-9}\cmidrule{11-14}\cmidrule{16-16}      & $Q^{(1)}(x)$ & $Q^{(2)}(x)$ & $Q^{(3)}(x)$ &       & $Q^{(1)}(x)$ & $Q^{(2)}(x)$ & $Q^{(3)}(x)$ & $Q^{PAA}(x)$ &       & $Q^{(1)}(x)$ & $Q^{(2)}(x)$ & $Q^{(3)}(x)$ & $Q^{PAA}(x)$ &       & $Q^{T-opt}(x)$ \\
\midrule
20 & 28.37 & 17.43 & 11.31 &       & 34.52 & 37.65 & 40.18 & 24.25 &       & 29.69 & 39.52 & 39.96 & 29.07 &       & 16.25 \\
30 & 28.69 & 15.05 & 10.82 &       & 31.71 & 33.04 & 30.24 & 21.02 &       & 29.20 & 32.71 & 30.48 & 23.90 &       & 16.52 \\
40 & 27.78 & 13.64 & 10.23 &       & 29.17 & 29.17 & 28.25 & 19.68 &       & 27.69 & 29.42 & 28.66 & 22.46 &       & 16.36 \\
50 & 28.39 & 13.39 & 10.85 &       & 29.41 & 26.69 & 26.73 & 19.39 &       & 28.66 & 27.94 & 27.71 & 21.88 &       & 16.32 \\
60 & 28.00 & 13.61 & 11.50 &       & 28.86 & 25.39 & 27.29 & 19.41 &       & 27.28 & 27.13 & 29.76 & 21.74 &       & 16.54 \\
70 & 28.39 & 13.71 & 12.53 &       & 29.14 & 24.25 & 23.61 & 18.72 &       & 27.75 & 25.12 & 26.68 & 20.10 &       & 16.63 \\
80 & 28.20 & 13.83 & 13.04 &       & 28.86 & 23.05 & 22.77 & 18.36 &       & 29.35 & 23.67 & 22.35 & 19.28 &       & 16.43 \\
90 & 28.19 & 13.94 & 13.40 &       & 28.77 & 22.17 & 21.19 & 17.74 &       & 32.63 & 22.89 & 21.69 & 19.49 &       & 16.32 \\
100 & 27.73 & 14.18 & 13.67 &       & 28.23 & 21.64 & 20.62 & 17.50 &       & 31.85 & 21.86 & 20.61 &18.96 &       & 16.26 \\
200 & 27.82 & 15.04 & 15.00 &       & 28.09 & 18.87 & 18.10 & 16.45 &       & 31.05 & 18.88& 18.11 & 17.09 &       & 16.36 \\
300 & 28.87 & 15.39 & 15.51 &       & 29.07 & 18.02 & 17.41 & 16.11 &       & 29.51 & 18.05 & 17.41 &16.60 &       & 16.42 \\
400 & 29.17 & 15.62 & 15.73 &       & 29.37 & 17.63 & 17.13 & 16.00 &       & 30.87 & 17.52 & 17.07 &15.89 &       & 16.33 \\
500 & 29.77 & 15.75 & 15.83 &       & 29.88 & 17.36 & 16.98 & 15.92 &       & 29.63 & 17.38 & 17.00 &15.88 &       & 16.37 \\

\bottomrule
\end{tabular}%
    \caption{Performance Comparison for Training, Testing, and Out-of-Sample Sets}
    \label{tab:fdcomparisons}
\end{table}

\begin{remark}
The effectiveness of the PAA policy in reducing overfitting lies in its innovative approach to determining weights based on the testing set. When weights are determined exclusively from the training set, machine learning policies — such as neural networks, which are prone to overfitting — tend to dominate. In contrast, by leveraging the testing set to determine the weights — similar to the cross-validation technique in machine learning — the PAA policy effectively mitigates overfitting. This approach not only balances the influence of different strategies but also enhances the overall robustness and performance of the model, ensuring that no single approach disproportionately dominates. 
\end{remark}

\section{Empirical Study}  \label{sec:empirical}
	
In this section, we compare PAA with established model-based policies and widely-used data-driven policies, including those referenced in \cite{ban2019big}. Through empirical analysis, we evaluate the PAA's performance in solving a real-life nurse staffing problem in a hospital. This problem requires determining the staffing level $Q$ for the next period, balancing the underage cost (e.g., increased mortality, reputational and legal damage) and the overage cost (e.g., higher total salary cost, unnecessary exposure risks). 

\begin{figure}[h!]
    \centering
    \begin{subfigure}[b]{0.48\textwidth}
        \centering
        \includegraphics[width=\textwidth]{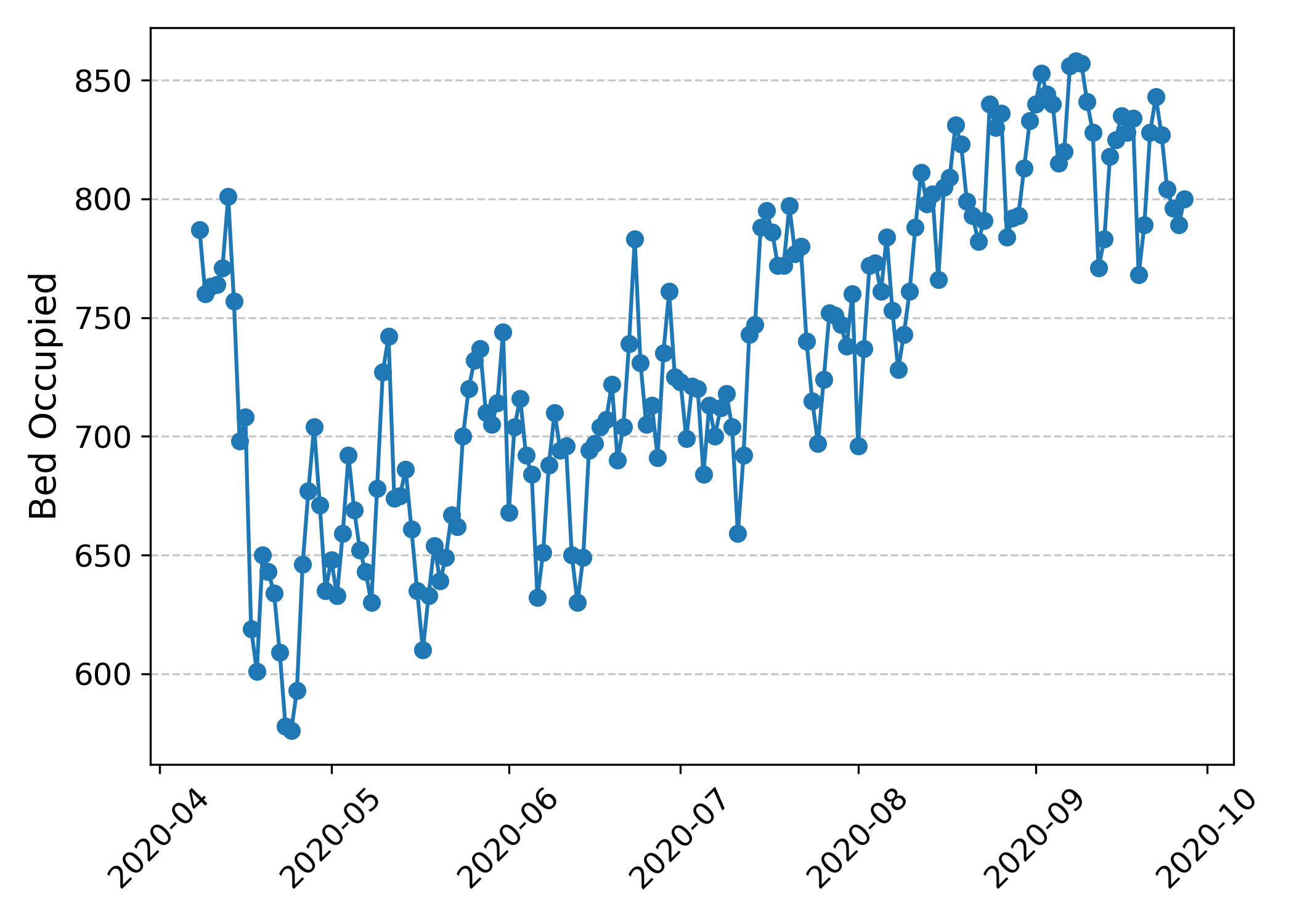}
        \caption{Time-series plot of bed occupancy}
        \label{fig:demand}
    \end{subfigure}
    \begin{subfigure}[b]{0.48\textwidth}
        \centering
        \includegraphics[width=\textwidth]{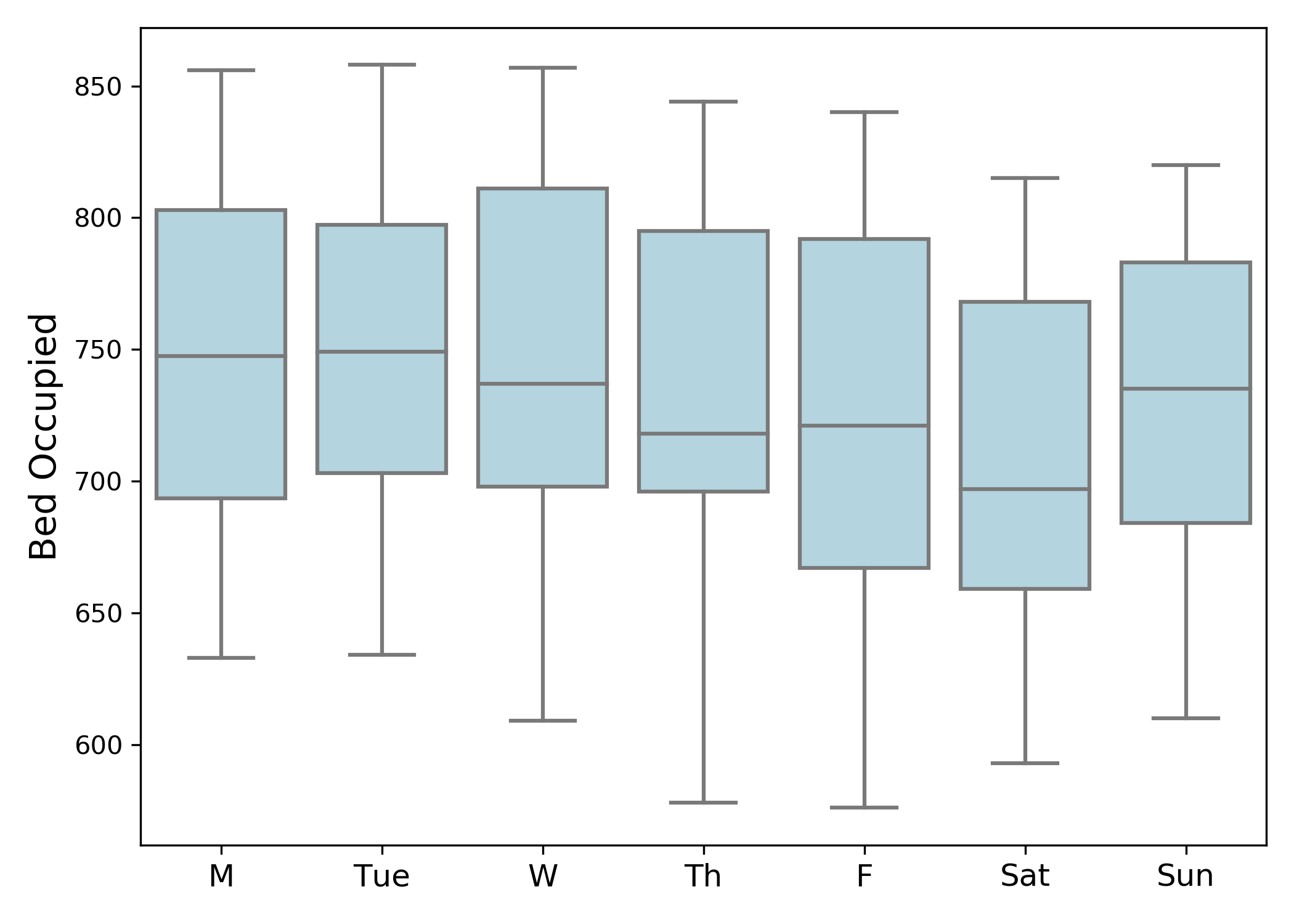}
        \caption{Box plot of bed occupancy by day}
        \label{fig:week}
    \end{subfigure}
    \caption{Hospital Bed Occupancy}
    \label{fig:combined}
\end{figure}

The dataset used originates from the NHS open dataset \cite{nhs2020}, which provides daily records of total bed occupancy for the London North West University Healthcare NHS Trust from April through October 2020. Figure~\ref{fig:demand} illustrates the time-series plot of bed occupancy, showing clear seasonality between mid-April and October, and stabilization from late-August onward. Figure~\ref{fig:week} provides a box plot of patient numbers by day of the week. Based on a typical nurse-to-patient ratio of 1:3, we estimate demand as the total number of patients divided by 3. Staffing recommendations do not require integer levels, as multi-skilled workers can be assigned part-time. Following \cite{ban2019big}, we set the underage cost as 2.5/3.5 and the overage cost as 1/3.5.

We apply covariate features inspired by \cite{ban2019big}, including the day of the week and the past demands over the previous three days. Additionally, we incorporate Operational Statistics (OS)—the sample average, minimum, maximum, and range of past demands over the previous 7 days to capture weekly periodicity. Using 183 observations from the dataset, we generate 177 usable data points after calculating OS features with a rolling 7-day window. To align with the apparent periodicity in Figure~\ref{fig:demand}, we retain 175 (a multiple of 7) data points, discarding the final two observations to ensure consistency. The dataset is split into training (70\%) and out-of-sample (30\%) sets, with training data used to compute candidate policies and PAA results, while out-of-sample data evaluates policy performance.

Overall, we consider 11 candidate policies in our study, consisting of three model-based (\( Q^{(1)}(\bm{x}) \), \( Q^{(2)}(\bm{x}) \), \( Q^{(3)}(\bm{x}) \)) and eight data-driven policies((\( Q^{(4)} \),\( Q^{(5)}(\bm{x}) \)–\( Q^{(11)}(\bm{x}) \)). 
Policy~\( Q^{(4)} \) is based on the Sample Average Approximation (SAA) approach in Section~\ref{subsec:newsvendordd} and does not consider features. This approach is used by \cite{levi2007provably} to address single-period and dynamic inventory problems with zero setup costs using nonparametric techniques. Except for 
\( Q^{(4)} \), all policies leverage the full set of features introduced above.
Policy~\( Q^{(6)}(\bm{x}) \) utilizes Quantile Regression (QR), a widely used statistical method for determining order quantities, selected here for its effectiveness in solving newsvendor problems. Policy~\( Q^{(7)}(\bm{x}) \)
implements Kernel Optimization (KO) of \cite{ban2019big} in Section~\ref{subsec:newsvendordd}.
Policies~\( Q^{(8)}(\bm{x}) \)-\( Q^{(11)}(\bm{x}) \) apply the Empirical Risk Minimization (ERM) approach of  \cite{ban2019big} in Section \ref{subsec:newsvendordd}.  
Details for the candidate policies are summarized in Table~\ref{tab:summary_of_policy}, including specific algorithms and parameters.
\begin{table}[!h]
\centering
\footnotesize
\renewcommand{\arraystretch}{1.2} 
\setlength{\tabcolsep}{8pt} 
\begin{tabular}{cll}
\toprule
\textbf{Abbreviation} & \textbf{Description} & \textbf{Parameter}  \\ 
\midrule
$Q^{(1)}(\bm{x})$ & ETO with Linear Regression & None  \\ 
$Q^{(2)}(\bm{x})$ & ETO with Polynomial Regression & Polynomial order = 2  \\ 
$Q^{(3)}(\bm{x})$ & ETO with ARIMA$(p,d,q)$ containing a Linear Trend & $p=1, d=0, q=0$  \\ 
$Q^{(4)}$ & SAA & None  \\ 
$Q^{(5)}(\bm{x})$ & First cluster then SAA & No. of clusters = 8 \\ 
$Q^{(6)}(\bm{x})$ & Quantile Regression & None \\ 
$Q^{(7)}(\bm{x})$ & KO with Gaussian Kernel & Bandwidth = 2  \\ 
$Q^{(8)}(\bm{x})$ & ERM & Polynomial order = 1  \\ 
$Q^{(9)}(\bm{x})$ & ERM & Polynomial order = 2  \\ 
$Q^{(10)}(\bm{x})$ & ERMreg1 & Polynomial order = 2  \\ 
$Q^{(11)}(\bm{x})$ & ERMreg2 & Polynomial order = 2  \\ 
\bottomrule
\end{tabular}
\caption{A Summary of the Candidate Policies Considered}
\label{tab:summary_of_policy}
\end{table}

First, we use all the 11 policies above as our candidate policies. Table~\ref{tab:cost_of_8_policy} provides a comparative analysis of various policies in terms of their average out-of-sample cost. With the lowest mean cost of 5.15, the PAA outperforms all other policies. 
The PAA shows a significant improvement of approximately 7.18\% over the best performing candidate policy $Q^{(11)}(\bm{x})$, demonstrating valuable cost savings. $Q^{(7)}(\bm{x})$, which demonstrates the best performance in the empirical results of \cite{ban2019big}, does not perform well on our data. The possible reasons for this discrepancy are as follows. First, the current dataset contains 119 samples with 14 features, which is insufficient for non-parametric approaches such as KO to cover the feature space effectively. Also, KO relies on kernel functions to weight local samples, and with an inadequate sample size, the kernel weights can become overly sparse. This sparsity often leads to predictions that deviate significantly from the actual values. 
Second, the current demand data exhibits clear linear trends that the KO approach struggles to capture. While KO performs local smoothing through kernel weighting, it lacks the ability to account for complex global patterns. Moreover, the KO approach tends to produce smooth prediction curves, which are more suitable for datasets with minor or continuous changes in demand, but it is less effective for datasets characterized by significant fluctuations.

\begin{table}[h!]
\centering
\footnotesize
\renewcommand{\arraystretch}{1.3} 
\setlength{\tabcolsep}{8pt} 
\begin{tabular}{lcccc}
\toprule
\textbf{Policies} & \textbf{25th Percentile} & \textbf{Median} & \textbf{75th Percentile} & \textbf{Mean} \\ 
\midrule
$Q^{(1)}(\bm{x})$ & 2.16 & 8.22 & 14.59 & 9.37 \\
$Q^{(2)}(\bm{x})$ & 2.72 & 5.34 & 11.39 & 6.98 \\
$Q^{(3)}(\bm{x})$ & 1.88 & 7.27 & 13.42 & 8.67 \\
$Q^{(4)}$ & 16.84 & 33.26 & 49.34 & 32.91 \\
$Q^{(5)}(\bm{x})$ & 8.93 & 24.29 & 38.93 & 24.83 \\
$Q^{(6)}(\bm{x})$ & 3.94 & 6.26 & 9.24 & 6.84 \\
$Q^{(7)}(\bm{x})$ & 14.29 & 30.36 & 41.43 & 28.37 \\
$Q^{(8)}(\bm{x})$ & 3.22 & 5.56 & 8.23 & 6.00 \\
$Q^{(9)}(\bm{x})$ & 2.65 & 4.67 & 7.41 & 5.66  \\
$Q^{(10)}(\bm{x})$ & 2.42 & 4.55 & 8.47 & 5.70 \\
$Q^{(11)}(\bm{x})$ & 2.04 & 3.96 & 8.12 & 5.52 \\
$Q^{PAA}(\bm{x})$ & 2.27 & 3.86 & 6.49 & 5.15 \\
\bottomrule
\end{tabular}
\caption{Comparison of PAA and Candidate Policies on Average Out-of-Sample Cost}
\label{tab:cost_of_8_policy}
\end{table}

\begin{figure}[h!]
    \centering
    \begin{subfigure}[b]{0.48\textwidth}
        \centering
        \includegraphics[width=\textwidth]{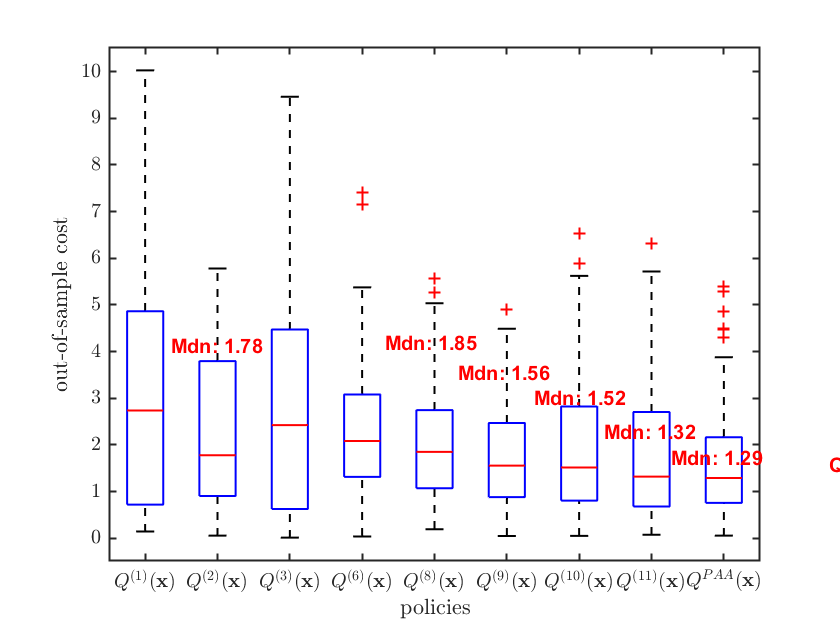} 
        \caption{All Policies}
        \label{fig:cost8:a}
    \end{subfigure}
    \hfill
    \begin{subfigure}[b]{0.48\textwidth}
        \centering
        \includegraphics[width=\textwidth]{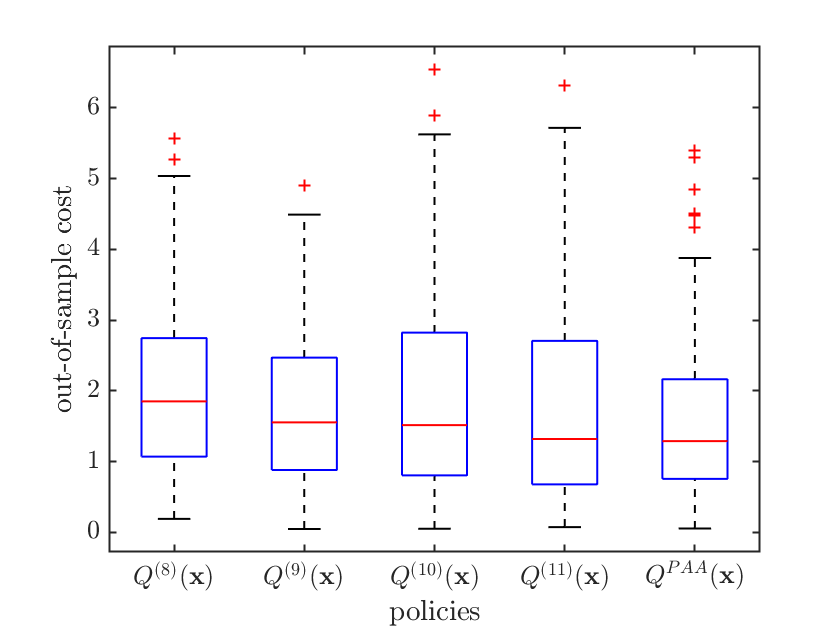} 
        \caption{ERM Versus PAA}
        \label{fig:cost8:b}    
    \end{subfigure}
    \caption{Box Plots of Out-of-Sample Cost}
    \label{fig:cost8}
\end{figure}

Figure~\ref{fig:cost8:a} shows a box plot of out-of-sample costs for all candidate policies, excluding policies $Q^{(4)},Q^{(5)}(\bm{x})$ and $Q^{(7)}(\bm{x})$. This exclusion is due to their significantly higher quantiles, which distort the display range of the other policies and make their performance appear indistinguishable. As shown in the figure, the PAA policy demonstrates superior performance, achieving the lowest median cost of 1.29 among all the evaluated policies. 
Figure~\ref{fig:cost8:b} shows a comparison among the \( Q^{(8)}(\bm{x}) \)-\( Q^{(11)}(\bm{x}) \) and PAA policies. These four are selected because they represent the policies with the lowest median costs and tighter distributions, as observed in Figure~\ref{fig:cost8:a}. In the more focused analysis presented in Figure~\ref{fig:cost8:b}, the PAA policy's advantages — its 
lowest median cost and smallest interquartile range — are highlighted even more clearly, further emphasizing its efficiency and robustness compared to the ERM-based policies \( Q^{(8)}(\bm{x}) \)-\( Q^{(11)}(\bm{x}) \).

\begin{figure}[htbp]
    \centering
    \includegraphics[width=0.7\textwidth]{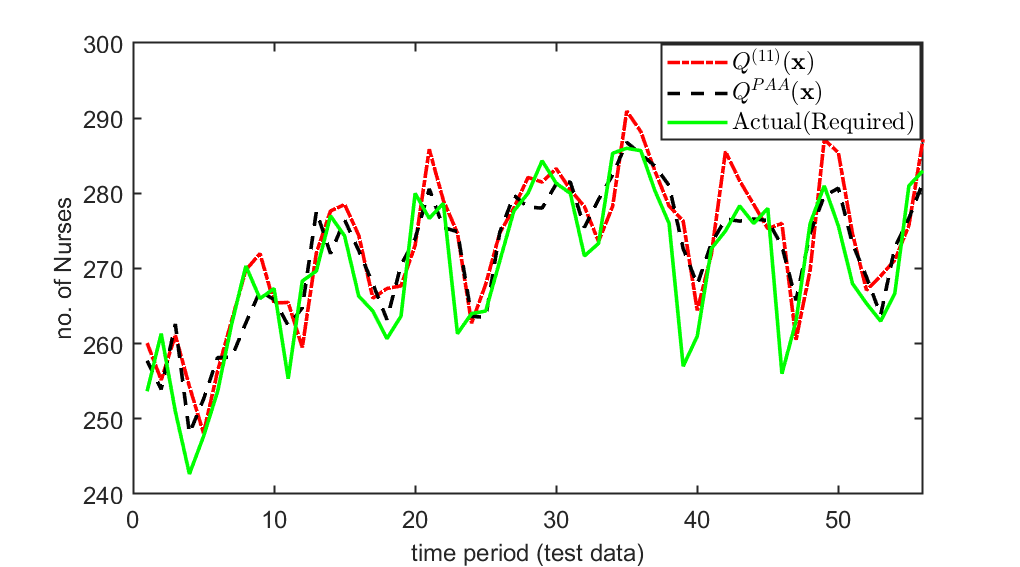}
    \caption{Actual Staffing Demand Versus Staffing Levels Predicted by ERM (Best Candidate Policy) and PAA on Out-of-Sample Set}
    \label{fig:order8}
\end{figure}

We further investigate the staffing decisions recommended by the PAA, as shown in Figure~\ref{fig:order8}. For comparison, the figure also includes the staffing levels predicted by the best candidate policy, \(Q^{(11)}(\bm{x})\), alongside the actual required number of nurses in the out-of-sample dataset. 
The actual number of nurses required over time (the thick green line) exhibits significant fluctuations, reflecting varying demand across the time periods. In addition to these fluctuations, the actual demand also shows a gradual upward trend, indicating that the overall demand for nurses is increasing over time. This upward trend might  be influenced by the worsening COVID-19 pandemic.
\( Q^{(11)}(\bm{x}) \) generally overestimates the number of nurses required. It has a higher mean absolute percentage deviation from actual (MAPD)  of \( 1.89\%\), whereas \( Q^{PAA}(\bm{x}) \) aligns better with the actual demand, achieving a lower MAPD of \( 1.62\%\). Thus, the PAA  demonstrates higher accuracy and more stable predictions over the time periods shown.

When selecting subsets of 2 to 10 policies as candidates from the 11 available, different PAA outcomes can be obtained. Table~\ref{tab:summary} provides a summary of the mean, median, 25th percentile  and 75th percentile out-of-sample costs for the optimal PAA as the number of candidate policies varies. From a theoretical perspective, according to Proposition~\ref{fbprop:PAA converges to the optimal}, when the data size is sufficiently large, the PAA converges to an optimal solution in the linear subspace spanned by the candidate policies. Therefore, increasing the number of candidate policies should theoretically improve the performance of the PAA. However, in the case of limited samples, this improvement is not guaranteed. Regardless of the number of selected policies, PAA consistently outperforms the standalone optimal policy \( Q^{(11)}(\bm{x}) \). Interestingly, the PAA with fewer than 11 candidate policies does not necessarily perform worse than with all 11 policies. For example, the lowest average out-of-sample cost is achieved when selecting 3 candidate policies, specifically \( Q^{(2)}(\bm{x}) \), \( Q^{(4)} \), \( Q^{(10)}(\bm{x}) \).  This finding illustrates that simply adding more candidate policies does not always result in improved performance.

Remarkably, not all strongly performing combinations include the best candidate policy $Q^{(11)}(\bm{x})$. Even when the best candidate policy \( Q^{(11)}(\bm{x}) \) is excluded, PAA can still outperform \( Q^{(11)}(\bm{x}) \) by leveraging policy \( Q^{(8)}(\bm{x}) \) which also has good performance. For example, when \( Q^{(7)}(\bm{x}) \) and the data-driven policy \( Q^{(8)}(\bm{x}) \) are selected as candidates, the average out-of-sample cost of PAA is \( 5.14 \). This result highlights a critical insight: the PAA does not rely solely on the inclusion of the best candidate policy but instead effectively utilizes the diversity and complementary strengths of other policies. This demonstrates the robustness of PAA and its ability to achieve superior performance through optimized policy combinations, even in the absence of the optimal candidate policy.
	
\begin{table}[!h]
\centering
\footnotesize
\renewcommand{\arraystretch}{1.3} 
\setlength{\tabcolsep}{8pt} 
\begin{tabular}{ccccccc}
\toprule
\textbf{Policies} & \textbf{Selected policies}  & \textbf{25th} & \textbf{Median} & \textbf{Mean} & \textbf{75th}&\textbf{Imp. vs. $\mathbf Q^{(11)}(\bm{x})$ (\%) }\\ 
\midrule
\multirow{7}{*}{$Q^{PAA}(\bm{x})$} 
 & \{7-8\} & 1.90 & 4.22 & 5.14 & 7.15&6.84\\
 & \{6-8\} & 2.18 & 4.41 & 5.06 & 6.80&8.32\\
 & \{4,6-8\} & 2.05&	4.21& 5.03 & 6.81&8.79\\
 & \{4,6-8,11\} & 2.19 & 4.33 & 5.06 & 6.84&8.34\\
 & \{2,3,8-11\} & 2.17 & 4.30 & 5.07 & 6.78 &8.19\\
 & \{1,2,,6,8-11\} & 2.42 & 4.69 & 5.09 & 6.92&8.46\\
 & \{2,4,6,7-11\} & 2.02 & 4.27 & 5.15 & 6.89&6.52 \\
 & \{1,3,5,6-11\} & 2.29 & 3.93 & 5.15 & 6.52&6.68\\
  & \{2-11\} & 2.16 & 4.02 & 5.10 & 6.65&6.96\\
  & \{1-11\} & 2.27 & 3.86 & 5.15 & 6.49&6.76\\
\bottomrule
\end{tabular}
\caption{PAA with Different Candidate Policies }
\label{tab:summary}
\end{table}

\section{Extensions}
\label{sec:extensions}

In this section, we briefly describe several extensions of the PAA that can improve decision-making for applications of stochastic optimization. Whereas our models above focus on the minimization of expected cost, tail risk is a concern for decision makers in various applications, for example in finance. We observe that the models presented above can be extended quite simply to incorporate standard measures of risk, including variance, value at risk and conditional value at risk. Using an appropriate weight adjustment, different attitudes towards risk can be modeled using the PAA.

A second extension is the use of stratified sampling among candidate policies, in order to balance the influence of different types of models. For example, we may have 80\% of candidate policies that are model-based and 20\% that are data-driven. However, for reasons including the need to provide a balanced justification for decisions, we may need both types of approach to have equal influence in decision making. To achieve this, we can layer the candidate policies by category and select an equal number from each category for final consideration. More generally, this methodology can support the empirical work of \cite{balakrishnan2024} on the integration of algorithmic recommendations and human decision making.

A final extension involves modeling the weights on candidate polices as functions of covariates instead of as scalar constants. In multi-period decision-making applications, this enables us to adjust scalar weights in discrete time period by period, but model inter-period effects in continuous time for greater modeling precision.

\section{Concluding Remarks} \label{sec:conclusions}
	
This work proposes a policy averaging approach (PAA) for stochastic optimization problems in general. The motivation for the PAA is to overcome the weaknesses of both the traditional model-based and data-driven approaches, but also to take advantage of their strengths. This is achieved using a variety of techniques to reduce variance and diversify among policies. The PAA collects various candidate policy recommendations derived from the modeling of problem data, and uses cross-validation to adjust them,  and optimization over weights to minimize the objective of the problem. We demonstrate the use of the PAA for the classical newsvendor problem, under two alternative models of random demand. Under i.i.d. demand, the optimization of weights is performed over demand quantiles. Under feature-based demand, however, the optimization of weights is performed over functions of demand features. We prove that the PAA recommends improved decisions compared to traditional model-based and data-driven approaches. An extensive simulation study further demonstrates that substantial improvement is achieved. We also conduct an empirical study using a real-world data set, again demonstrating a substantial improvement from the use of the PAA. Extensions to consider tail risk, and the use of stratified sampling to balance the influence of different approaches, are discussed.

From the perspective of managerial decision making, the PAA provides intuitive support for policies based on its underlying use of models. At the same time, it takes advantage of the large data sets that are increasingly available, to develop robust recommendations. The combination of advanced modeling, variance reduction, optimized policy diversification and where necessary stratified sampling provides strong justification for the policy recommendations of the PAA. More philosophically, the concept of implemented collective wisdom is a potentially persuasive one to support managerial recommendations.

For future research beyond the classical newsvendor problem, we recommend the application of the PAA to a variety of stochastic optimization problems. First, there are various extensions of the simple newsvendor problem to include backordering, or multiple periods with inventory carryover. Second, a natural extension is the consideration of multi-period inventory problems under stochastic demand, including lot size and reorder point systems. Other high value applications to which the PAA can be applied include the design of acceptance sampling plans in quality control, and the setting of warranties and of insurance premiums. Indeed, the PAA offers potential value in any stochastic optimization environment for which large data sets are available.

\bibliographystyle{apalike}
\bibliography{ref}

\newpage
\appendix
\section*{Online Appendix A: Proofs}
\subsection*{Appendix A1: Proof of Proposition \ref{iidprop:PAA is the best}}

Under the assumptions of i.i.d. and infinite sample size, the objective of $(PAA_{iid})$ converges to the real expected cost $\mathbb{E}[C(Q-d)]$. As $L\leq 0$ and $U\geq 1$, we can find that any candidate policy is an admissible policy in problem $(PAA_{iid})$. Due to the optimality of the PAA policy, it must have an objective that is no larger than any candidate policy.

\subsection*{Appendix A2: Proof of Proposition \ref{iidprop:PAA converges to the optimal}}
 The variance of the PAA policy is upper-bounded by:
 \begin{align*}
 \mbox{Var}(Q^{PAA}(\bm D_t)) &= \mbox{Var}\Big(\sum_{i=1}^m \omega_i Q^{(i)}(\bm D_t)\Big) \\
 &= \sum_{i=1}^m \omega_i^2 \mbox{Var}(Q^{(i)}(\bm D_t)) + 2\sum_{1\leq i<j \leq m} \omega_i\omega_j \mbox{Cov}(Q^{(i)}(\bm D_t),Q^{(j)}(\bm D_t)) \\
 &\leq \frac{1}{t} \sum_{i=1}^m \omega_i^2 M_i^2 + \frac{2}{t}\sum_{1\leq i<j \leq m} \omega_i\omega_j M_i M_j \\
 &\leq \frac{m^2 \cdot \max (L^2,U^2) \cdot \max_i M_i^2}{t}.
 \end{align*}
 As long as  \(L\) and \(U\) for data set $\bm D_t$ approach negative and positive infinity, respectively, at a rate slower than \(-\sqrt{t}\) and \(\sqrt{t}\), it follows that 
 $$\lim_{t \to \infty} \mbox{Var}(Q^{PAA}(\bm D_t)) = 0,$$ 
 which implies that the PAA policy converges to a constant.
 
 Moreover, the limit of the PAA policy is given by:
\[
\lim_{t \to \infty} Q^{PAA}(\bm{D}_t) = \sum_{i=1}^m \omega_i \bar{Q}^{(i)}.
\]
Given the assumption that \(\bar{Q}^{(i)} \neq \bar{Q}^{(j)}\) for some \(i \neq j\), the expectation of the PAA policy spans the entire space \(\mathbb{R}\), which must include the optimal order quantity \(Q^{T-opt}\). Therefore, due to the optimality of the PAA policy, the limit value must converge to \(Q^{T-opt}\). 

\subsection*{Appendix A3: Proof of Proposition \ref{iidprop:corr dis}}
We first derive the expected cost of a candidate policy \( Q^{(i)}(\bm{D}_t) \), for \( i = 1, 2 \), which follows a normal distribution \( N(\mu_i, \sigma_i^2) \), under the assumption that the demand is also normally distributed as \( N(\mu_d, \sigma_d^2) \). Furthermore, we assume that the demand and candidate policies are independent. Under this assumption, it is clear that:
\(
Q^{(i)}(\bm{D}_t) - d \sim N(\mu_i - \mu_d, \sigma_i^2 + \sigma_d^2).
\)

The expected cost is expressed by
\begin{align*}
    \E [C(Q^{(i)}(\bm D_t) - d)] = c_o\E[ \max(Q^{(i)}(\bm D_t) - d, 0)] + c_u\E [\max(d - Q^{(i)}(\bm D_t), 0)].
\end{align*}
Considering the overordering term, we have:
\begin{align*}
&\E[\max(Q^{(i)}(\bm D_t) - d, 0)] \\
=& \int_0^{\infty} (Q - d) \frac{1}{\sqrt{2 \pi (\sigma_i^2 + \sigma_d^2)}} e^{-\frac{\left(Q - d - (\mu_i - \mu_d)\right)^2}{2 (\sigma_i^2 + \sigma_d^2)}} \, d(Q - d)\\
= & \sqrt{\sigma_i^2 + \sigma_d^2} \int_{\frac{\mu_d - \mu_i}{\sqrt{\sigma_i^2+ \sigma_d^2}}}^{\infty} 
\frac{z}{\sqrt{2 \pi}} e^{-\frac{z^2}{2}}  dz + (\mu_i - \mu_d) \int_{\frac{\mu_d - \mu_i}{\sqrt{\sigma_i^2+ \sigma_d^2}}}^{\infty} 
\frac{1}{\sqrt{2 \pi}} e^{-\frac{z^2}{2}}  dz\\
= & \frac{\sqrt{\sigma_i^2 + \sigma_d^2}}{\sqrt{2\pi}} e^{-\frac{(\mu_i - \mu_d)^2}{2(\sigma_i^2 + \sigma_d^2)}} + (\mu_i - \mu_d) \Big[1 - \Phi\Big(-\frac{\mu_i-\mu_d}{\sqrt{\sigma_i^2 + \sigma_d^2}}\Big)\Big].
\end{align*}
Similarly, the underordering term is:
\begin{align*}
& \E[\max(d-Q^{(i)}(\bm D_t), 0)] \\
=& -\int_{-\infty}^{0} (Q - d) \frac{1}{\sqrt{2 \pi (\sigma_i^2 + \sigma_d^2)}} e^{-\frac{\left(Q - d - (\mu_i - \mu_d)\right)^2}{2 (\sigma_i^2 + \sigma_d^2)}} \, d(Q - d)\\
= & -\sqrt{\sigma_i^2 + \sigma_d^2} \int_{-\infty}^{\frac{\mu_d - \mu_i}{\sqrt{\sigma_i^2+ \sigma_d^2}}} 
\frac{z}{\sqrt{2 \pi}} e^{-\frac{z^2}{2}}  dz - (\mu_i - \mu_d) \int_{-\infty}^{\frac{\mu_d - \mu_i}{\sqrt{\sigma_i^2+ \sigma_d^2}}} 
\frac{1}{\sqrt{2 \pi}} e^{-\frac{z^2}{2}}  dz\\
= & \frac{\sqrt{\sigma_i^2 + \sigma_d^2}}{\sqrt{2\pi}} e^{-\frac{(\mu_i - \mu_d)^2}{2(\sigma_i^2 + \sigma_d^2)}} - (\mu_i - \mu_d) \Big[1 - \Phi\Big(\frac{\mu_i-\mu_d}{\sqrt{\sigma_i^2 + \sigma_d^2}}\Big)\Big].
\end{align*}
Combining these results, we obtain:
\begin{align*}
  \mathbb{E}[C(Q^{(i)}(\bm{D}_t) - d)] = ~& (c_o + c_u) \frac{\sqrt{\sigma_i^2 + \sigma_d^2}}{\sqrt{2\pi}} e^{-\frac{(\mu_i - \mu_d)^2}{2(\sigma_i^2 + \sigma_d^2)}} + (\mu_i - \mu_d) (c_o - c_u) \\
  & + (\mu_i - \mu_d) \Big[ c_u \Phi\Big(\frac{\mu_i - \mu_d}{\sqrt{\sigma_i^2 + \sigma_d^2}} \Big) - c_o \Phi\Big( \frac{\mu_d - \mu_i}{\sqrt{\sigma_i^2 + \sigma_d^2}} \Big) \Big],
\end{align*}
which is denoted as  \(J(\mu_i, \sigma_i^2).\)

Under the distribution assumption in the proposition, we have \( \mathbb{E}[Q(\bm{D}_t)] = \mu \) and the variance of \( Q(\bm{D}_t) \) is given by 
\(
\mbox{Var}(Q(\bm{D}_t)) = (\omega_1^2 + (1 - \omega_1)^2 + 2\rho \omega_1(1 - \omega_1)) \sigma^2,
\)
where \( Q(\bm{D}_t) = \omega_1 Q^{(1)}(\bm{D}_t) + (1 - \omega_1) Q^{(2)}(\bm{D}_t) \). 
The first-order derivative of the expected cost with respect to \( \omega_1 \) is:
\begin{align*}
 &\frac{d \E[C (Q(\bm D_t)-d)]}{d \omega_1} \\
 = &\, \frac{d \E[C (Q(\bm D_t)-d)]}{d \sqrt{\operatorname{Var}(Q(\bm D_t)) + \sigma_d^2}} \frac{d\sqrt{\operatorname{Var}(Q(\bm D_t)) + \sigma_d^2}}{d \omega_1}\\
  = &\,  
  \Bigg[\frac{c_o+c_u}{\sqrt{2 \pi}} e^{-\frac{(\mu-\mu_d)^2}{2(\operatorname{Var}(Q) + \sigma_d^2)}} + \frac{(c_o+c_u)(\mu-\mu_d)^2}{\sqrt{2 \pi}(\mbox{Var}(Q(\bm D_t))+\sigma_d^2)} e^{-\frac{(\mu-\mu_d)^2}{2(\operatorname{Var}(Q) + \sigma_d^2)}} \\
  & - \frac{(c_o+c_u)(\mu-\mu_d)^2}{\sqrt{2 \pi}(\mbox{Var}(Q(\bm D_t))+\sigma_d^2)} e^{-\frac{(\mu-\mu_d)^2}{2(\operatorname{Var}(Q) + \sigma_d^2)}}  \Bigg] 
  \frac{\sigma^2(1-\rho)(2 \omega_1-1)}{\sqrt{(\operatorname{Var}(Q) + \sigma_d^2)}}\\
  = &\,  
  \frac{c_o+c_u}{\sqrt{2 \pi}} e^{-\frac{(\mu-\mu_d)^2}{2(\operatorname{Var}(Q) + \sigma_d^2)}} 
  \frac{\sigma^2(1-\rho)(2 \omega_1-1)}{\sqrt{(\operatorname{Var}(Q) + \sigma_d^2)}}.
 \end{align*}
Then, the first order condition implies that the optimal $\omega_1^*$ is $\frac{1}{2}$. Hence, the variance of PAA policy is \(Var(Q^{PAA}(\bm D_t))=\sigma^2 (1 + \rho) / 2\), and
the improvement over the individual candidate policies is:
\[
J(\mu, \sigma^2) - J(\mu,  (1 + \rho) / 2\sigma^2). 
\]

\subsection*{Appendix A4: Proof of Proposition \ref{iidprop:mean dis}}
Under the distribution assumption in the proposition, we have \( \mbox{Var}(Q(\bm{D}_t)) = \sigma^2 \) and the mean of \( Q(\bm{D}_t) \) is given by 
\(
\E[Q(\bm{D}_t)] = \omega_1\mu_1+(1-\omega_1)\mu_2,
\)
where \( Q(\bm{D}_t) = \omega_1 Q^{(1)}(\bm{D}_t) + (1 - \omega_1) Q^{(2)}(\bm{D}_t) \). 
The first-order derivative of the expected cost with respect to \( \E[Q(\bm D_t)] -d  \) is:
\begin{align*}
 & \frac{d \E[C (Q(\bm D_t)-d)]}{d (\E[Q(\bm D_t)] - \mu_d)} \\
 = & \, -\frac{(c_o+c_u)(\E[Q(\bm D_t)]-\mu_d)}{\sqrt{2\pi(\sigma^2+\sigma_d^2)}}e^{-\frac{(\E[Q(\bm D_t)]-\mu_d)^2}{2(\sigma^2+\sigma_d^2)}}+(c_o-c_u) + c_u \Phi\Big(\frac{\mu_i - \mu_d}{\sqrt{\sigma_i^2 + \sigma_d^2}} \Big) - c_o \Phi\Big( \frac{\mu_d - \mu_i}{\sqrt{\sigma_i^2 + \sigma_d^2}} \Big)\\
 & + \frac{(c_o+c_u)(\E[Q(\bm D_t)]-\mu_d)}{\sqrt{2\pi(\sigma^2+\sigma_d^2)}}e^{-\frac{(\E[Q(\bm D_t)]-\mu_d)^2}{2(\sigma^2+\sigma_d^2)}}\\
 =& \, c_u \Phi\Big(\frac{\mu_i - \mu_d}{\sqrt{\sigma_i^2 + \sigma_d^2}} \Big) - c_o \Phi\Big( \frac{\mu_d - \mu_i}{\sqrt{\sigma_i^2 + \sigma_d^2}} \Big)+c_o-c_u.
 \end{align*}
Letting $\frac{d \E[C (Q(\bm D_t)-d)]}{d (\E[Q(\bm D_t)] - \mu_d)}=0$, we obtain 
\[
E[Q^{PAA}(\bm D_t)] - \mu_d=\sqrt{\sigma^2+\sigma_d^2}\Phi^{-1}(\frac{c_u}{c_o+c_u}),
\]
which implies
\[
\E[Q^{PAA}(\bm D_t)] = \mu_d+\sqrt{\sigma^2+\sigma_d^2}\Phi^{-1}(\frac{c_u}{c_o+c_u}).
\]

It is clear that the improvement in the expected cost of the PAA policy over the candidate policies is given by 
\(
J\left(\E[Q^{PAA}(\bm{D}_t)], \sigma^2\right) - J(\mu_1, \sigma^2) \geq 0
\)
and 
\(
J\left(\E[Q^{PAA}(\bm{D}_t)], \sigma^2\right) - J(\mu_2,\sigma^2) \geq 0,
\)
respectively.
	
\subsection*{Appendix A5: Proof of Proposition \ref{iidprop:impro from dis}}
For the PAA policy, the objective function of $(PAA_{iid})$ is given by:
 \begin{align*}
 & \frac{c_o}{t} \sum_{j=1}^t \max (Q(\bm{D}_{\cancel{j}}) - d_j, 0) + \frac{c_u}{t} \sum_{j=1}^t \max (d_j - Q(\bm{D}_{\cancel{j}}), 0) \\
 = & \frac{c_o}{t} \sum_{j=1}^t \max \left(Q^{(1)}(\bm{D}_{\cancel{j}}) + \omega_2 \left(Q^{(2)}(\bm{D}_{\cancel{j}}) - Q^{(1)}(\bm{D}_{\cancel{j}})\right) - d_j, 0 \right) \\
 & + \frac{c_u}{t} \sum_{j=1}^t \max \left(d_j - Q^{(1)}(\bm{D}_{\cancel{j}}) - \omega_2 \left(Q^{(2)}(\bm{D}_{\cancel{j}}) - Q^{(1)}(\bm{D}_{\cancel{j}})\right), 0 \right).
 \end{align*}
 
 To consider the extreme case where the objective of the PAA policy can be minimized as much as possible, we analyze two scenarios:
 
 1. When \(c_o \geq c_u\):
 
 If \(Q^{(1)}(\bm{D}_{\cancel{j}}) + \omega_2 \left(Q^{(2)}(\bm{D}_{\cancel{j}}) - Q^{(1)}(\bm{D}_{\cancel{j}})\right) - d_j \geq 0\) and \(\omega_2 \left(Q^{(2)}(\bm{D}_{\cancel{j}}) - Q^{(1)}(\bm{D}_{\cancel{j}})\right) \leq 0\) for all \(j\), then the objective becomes
 \begin{align*}
 & \frac{c_o}{t} \sum_{j=1}^t \max (Q(\bm{D}_{\cancel{j}}) - d_j, 0) + \frac{c_u}{t} \sum_{j=1}^t \max (d_j - Q(\bm{D}_{\cancel{j}}), 0) \\
 =&\frac{c_o}{t} \sum_{j=1}^t \max \left(Q^{(1)}(\bm{D}_{\cancel{j}}) - d_j, 0 \right) + \frac{c_u}{t} \sum_{j=1}^t \max \left(d_j - Q^{(1)}(\bm{D}_{\cancel{j}}), 0 \right) \\
 & - \frac{c_o}{t} \sum_{j=1}^t \left|\omega_2 \left(Q^{(1)}(\bm{D}_{\cancel{j}}) - Q^{(2)}(\bm{D}_{\cancel{j}})\right)\right| \\
 \geq & \frac{c_o}{t} \sum_{j=1}^t \max \left(Q^{(1)}(\bm{D}_{\cancel{j}}) - d_j, 0 \right) + \frac{c_u}{t} \sum_{j=1}^t \max \left(d_j - Q^{(1)}(\bm{D}_{\cancel{j}}), 0 \right) - c_o \max (|L|,U) \delta,
 \end{align*}
where the first equality holds due to $\frac{c_u}{t} \sum_{j=1}^t \max \left(d_j - Q^{(1)}(\bm{D}_{\cancel{j}}), 0 \right)=0$, the second inequality holds due to the definition of $\ell_1$ distance,  
\(
\frac{1}{t} \sum_{j=1}^t \left| Q^{(1)}(\bm{D}_{\cancel{j}}) - Q^{(2)}(\bm{D}_{\cancel{j}}) \right| = \delta.
\)
 
 2. When \(c_u \geq c_o\), the improvement bound becomes \(c_u \max \{|L|,U\} \delta\). 
 
Thus, the improvement of the PAA policy over the first candidate policy is bounded by \(\max\{c_o, c_u\} \max \{|L|,U\} \delta\). Similarly, the same upper bound applies to the second candidate policy.
 
\subsection*{Appendix A6: Proof of Proposition \ref{fbprop:PAA is the best}}
The proof of this proposition follows the same reasoning as the proof of Proposition~\ref{iidprop:PAA is the best}.

\subsection*{Appendix A7: Proof of Proposition \ref{fbprop:PAA converges to the optimal}}
The variance of the coefficient $q^*_k(\bm D_t)$ in PAA policy is upper-bounded by:
    \begin{align*}
		\mbox{Var}(q_k^*(\bm D_t)) &= \mbox{Var}\Big(\sum_{i=1}^m \omega_i q^{(i)}_k(\bm D_t)\Big) \\
        & = \sum_{i=1} \omega_i^2 \mbox{Var}(q^{(i)}_k(\bm D_t)) +2  \sum_{1\leq i< j\leq m} \omega_i\omega_j \mbox{Cov}(q^{(i)}_k(\bm D_t),q^{(j)}_k(\bm D_t))\\
    	&\leq \frac{1}{t} \sum_{i=1}^m \omega_i^2 M_{ik}^2 + \frac{2}{t}\sum_{1\leq i<j \leq m} \omega_i\omega_j M_{ik} M_{jk} \\
 &\leq \frac{m^2 \cdot \max (L^2,U^2) \cdot \max_i M_{ik}^2}{t},
	\end{align*}
where \( M_{ik}\) is the \(k\)th element of \(\bm M_i\).
    
As long as \(L\) and \(U\) for the data set \(\bm{D}_t\) approach negative and positive infinity, respectively, at rates slower than \(-\sqrt{t}\) and \(\sqrt{t}\) , it follows that 
	\[
    \lim_{t \to \infty} \mbox{Var}(q_k^*(\bm D_t)) = 0,
       \]
which implies that the coefficients of the PAA policy converges to a constant vector.

Moreover, the coefficients of the PAA policy converge to 
	\begin{align*}
		\lim_{t\to\infty}\bm q^*(\bm D_t) = \sum_{i=1}^m \omega_i \bar{\bm q}^{(i)}.
	\end{align*}
Under the assumption that \(\{\bar{\bm{q}}^{(1)}, \dots, \bar{\bm{q}}^{(m)}\}\) form a system of vectors containing \(p+1\) independent vectors, the admissible PAA policy covers the entire function space of polynomials of degree $p$. The optimality of the PAA policy implies that it converges to the theoretical optimal order quantity function as \(t \to \infty\).

\subsection*{Appendix A8: Proof of Proposition \ref{fbprop:impro from dis}}
The proof of this proposition follows the same reasoning as the proof of Proposition~\ref{iidprop:impro from dis}.
 
\subsection*{Appendix A9: Proof of Proposition \ref{fbprop:overfitting}}
According to the proof of Proposition \ref{iidprop:corr dis}, the optimal weight on the first candidate policy is $\omega_1^*=\frac{1}{2}$. 

Then, we have  
\begin{align*}
& \operatorname{Var}(Q^{PAA}(x;\bm D_t)) - \operatorname{Var}\Big(\frac{1}{2} \beta^{(1)}(x;\bm D_t) + \frac{1}{2}\beta^{(2)}(x;\bm D_t)\Big) - \sigma^2 + \sigma_{\beta}^2\\
= &\,\frac{(1+\rho)\sigma^2}{2} - \frac{(1+\rho_{\beta})\sigma_{\beta}^2}{2} - \sigma^2 + \sigma_{\beta}^2\\
=& \, \frac{\rho \sigma^2 - \rho_\beta \sigma_\beta^2}{2} - \frac{\sigma^2-\sigma_\beta^2}{2}\\
<&\, 0,
\end{align*}
which implies the inequality in proposition holds. 

\end{document}